\documentclass[11pt]{article}
\usepackage{a4,bm,epsfig,booktabs}
\usepackage{setspace}
\newcommand{\paramd}{d}
\newcommand{\parame}{e}
\newcommand{\fktg}{G}

\newcommand{\Abinv}{\Ab_{\mathrm{inv}}}
\newcommand{\neueseite}{\newpage}
\newcommand{\nonicht}[1]{\textbf{Das ist hier noch unbesetzt.}}
\newcommand{\Bigeinschr}[1]{\Big|_{#1}}
\usepackage{diagrams}
\newcommand{\ogkl}[1]{\lceil#1\rceil}
\newcommand{\ddfrac}[2]{\frac{\dd#1}{\dd#2}}
\newcommand{\AP}{{\mathrm{AP}}}
\newcommand{\grav}{\text{grav}}
\newcommand{\cosm}{\text{cosm}}
\newcommand{\paraml}{\lambda}
\newcommand{\zspeed}{\nu}
\newcommand{\faf}{\alpha}
\newcommand{\fbf}{\beta}
\newcommand{\ffa}{\rho_0}
\newcommand{\ffb}{\rho_1}
\newcommand{\ffd}{\sigma_{11}}
\newcommand{\ffdd}{\sigma_{10}}
\newcommand{\ffe}{\sigma_{00}}
\newcommand{\fff}{f}
\newcommand{\ffg}{g}
\newcommand{\intsqrtf}{\xi}
\newcommand{\grenzfaf}{\faf_\infty}
\newcommand{\restfaf}{\faf_0}

\newarrow{auf}|---{->}
\newarrow{nach}----{->}
\newarrow{ident}33333
\newarrow{aufsurj}|---{->>}
\newarrow{nachsurj}----{->>}
\newarrow{einbett}C---{->}
\newarrow{strichel}{}{dash}{}{dash}{->}
\newarrow{impliz}===={=>}
\newarrow{fillauf}....{->}
\newlength{\CDhoehe}                  
\newlength{\CDgap}                    
\newcommand{\vc}{{\mathbf{c}}}

\makeatletter
\newcounter{tab}
\newcommand{\fusszeile}{}
                        {\par\noindent\fusszeile
                         \end{center}}
\makeatother
\usepackage{ifthen,array}
\newcommand{\ams}{\usepackage{amsfonts,amssymb,amsmath}}

\ams
\allowdisplaybreaks[3]
\newlength{\textwidthorig}
\newlength{\oddsidemarginorig}
\newlength{\textheightorig}
\newlength{\topmarginorig}
\def\seitenlaengenabsolut#1 #2 #3 #4 {\setlength{\textwidth}{#1}
                                      \setlength{\oddsidemargin}{#2}
                                      \setlength{\textheight}{#3}
                                      \setlength{\topmargin}{#4}}
\def\seitenlaengenrelzustandard#1 #2 #3 #4 {\setlength{\textwidth}{\textwidthorig+#1}
                                            \setlength{\oddsidemargin}{\oddsidemarginorig+#2}
                                            \setlength{\textheight}{\textheightorig+#3}
                                            \setlength{\topmargin}{\topmarginorig+#4}}
\def\seitenlaengenrelzuvorher#1 #2 #3 #4 {\addtolength{\textwidth}{#1}
                                          \addtolength{\oddsidemargin}{#2}
                                          \addtolength{\textheight}{#3}
                                          \addtolength{\topmargin}{#4}}
\newcommand{\standardseite}{\seitenlaengenrelzuvorher2.2cm -0.8cm 1.8cm -1.5cm }   %
\standardseite
\newcommand{\leerezeile}{\vspace{2ex}}
\newlength{\laengespatium}

\newcommand{\nach}{\longrightarrow}      

\newcommand{\auf}{\longmapsto}           
\newcommand{\txtauf}[1]{\auf}            
\newcommand{\impliz}{\Longrightarrow}    
\newcommand{\aequ}{\Longleftrightarrow}  
 
\newcommand{\invimpliz}{\Longleftarrow}  
\newcommand{\gegen}{\rightarrow}         
\newcommand{\iso}{\cong}                 
\newcommand{\ident}{\equiv}              
\newcommand{\teilmenge}{\subseteq}       
\newcommand{\obermenge}{\supseteq}       
\newcommand{\aeqrel}{\sim}               

\newcommand{\fueralle}{\hspace{1.7em}\forall}
\newcommand{\leeremenge}{\varnothing}    
\newcommand{\kreuz}{\times}              

\newcommand{\einschr}[1]{{}\arrowvert_{#1}}      

\newcommand{\betraganpass}[1]%
           {\left| #1 \right|}           
\newcommand{\bigbetrag}[1]%
           {\bigl|{#1}\bigr|}            
\newcommand{\betrag}[1]%
           {|{#1}|}                      
\newcommand{\betragnichtanpass}[1]%
           {\mid #1 \mid}                
\newcommand{\norm}[1]%
           {{}{\parallel}#1{\parallel}{}}      
\newcommand{\erww}[1]%
           {\langle #1 \rangle}          
\newcommand{\skalprod}[2]%
           {\langle #1,#2 \rangle}       
\newcommand{\supnorm}[1]{{\norm{#1}_\infty}}        
\newcommand{\quer}{\overline}            
\newcommand{\dach}{\widehat}             
\newcommand{\inv}[1]{\frac{1}{#1}}       
\newcommand{\einhalb}{\inv{2}}           
\newcommand{\re}{\text{Re }}                           
\newcommand{\dd}{\text{d}}                             
\newcommand{\e}{\text{e}}                              
\newcommand{\I}{\text{i}}                              
\newcommand{\NULL}{\mathbf{0}}                         
\newcommand{\EINS}{{\boldsymbol{1}}}                   
\newcommand{\field}[1]{\mathbb{#1}}                    
\newcommand{\C}{{\field{C}}}                           
\newcommand{\N}{{\field{N}}}                           
\newcommand{\R}{{\field{R}}}                           
\newcommand{\Q}{{\field{Q}}}                           
\newcommand{\Z}{{\field{Z}}}                           
\newcommand{\boundfkt}{{\cal B}}                       
\newcommand{\rnkl}[2]{\raisebox{-0.4ex}{$#1$}%
\raisebox{-0.12ex}{{\large$\setminus$}}\,#2}   
\newcommand{\agb}{{\overline{{\cal A}/{\cal G}}}}      
\newcommand{\agbfact}[1][]{\text{$\agb/\!\aeqrel$}}    
\newcommand{\Ab}{{\overline{{\cal A}}}}                
\newcommand{\A}{{\cal A}}                              
\newcommand{\Gb}{{\overline{{\cal G}}}}                

\newcommand{\qa}{{\quer{A}}}                           

\newcommand{\holgr}{{\mathbf H}}                       
\newcommand{\bz}{{\mathbf B}}                          

\newcommand{\GR}{\Gamma}                               
\newcommand{\Pf}{{\cal P}}                             
\newcommand{\Haar}{{\text{Haar}}}                      
\newcommand{\LG}{{\mathbf{G}}}                         
\newcommand{\aeqrelzush}[1][]{\sim}                    
\newcommand{\alg}{\mathfrak{A}}                          
\newcommand{\blg}{\mathfrak{B}}                          
\newcommand{\malg}{{\cal M}(\alg)}                     

\newcommand{\nklza}[1][]{\ifthenelse{\equal{#1}{}}     
                                    {\rnkl{Z(\holgr_\qa)}{\LG}}        
                                   {\rnkl{Z(\holgr_{#1})}{\LG}}}       
\newcommand{\nkla}[1][]{\ifthenelse{\equal{#1}{}}      
                                    {\rnkl{\bz(\qa)}{\Gb}}        
                                    {\rnkl{\bz(#1)}{\Gb}}}       

\newcommand{\charakt}{\chi}                            

\newcommand{\YM}{{\text{YM}}}                          

\newcommand{\ymwirk}[1][]{\ifthenelse{\equal{#1}{}}{S_{\YM}}{S_{\YM,#1}}}

\newcommand{\bmat}{\begin{pmatrix}}
\newcommand{\emat}{\end{pmatrix}}

\newcommand{\ListNullAbstaende}{\setlength{\topsep}{0pt}%
                                \setlength{\parskip}{0pt}%
                                \setlength{\partopsep}{0pt}%
                                \setlength{\itemsep}{0pt}%
                                \setlength{\parsep}{0pt}}
\newcommand{\ListNurAnstrichAbstand}{\setlength{\topsep}{0pt}%
                                     \setlength{\parskip}{0pt}%
                                     \setlength{\partopsep}{0pt}%
                                     \setlength{\parsep}{0pt}}
\newenvironment{StandardListe}[2]%
               {\begin{list}%
                      {#1}%
                      {\settowidth{\leftmargin}{M#1}%
                       \settowidth{\labelwidth}{#1}%
                       \settowidth{\labelsep}{M}%
                       #2%
                      }%
                }%
               {\end{list}}%
\newenvironment{EinfachListe}[1]%
               {\begin{StandardListe}{#1}{\ListNullAbstaende}}%
               {\end{StandardListe}}%
               {\begin{StandardListe}{#1}{\ListNurAnstrichAbstand}}%
               {\end{StandardListe}}%
\newcommand{\labelsatz}[1]{#1}
\newcounter{listennr}                      %
\newlength{\hilfslaenge}
\newlength{\stdlabellaenge}
\newlength{\maximum}
\newcommand{\stdlabel}{}
\newcommand{\Maximum}{}
\newcommand{\iitem}[1][]{\ifthenelse{\equal{#1}{}}%
                           {\item \setlength{\hilfslaenge}{\stdlabellaenge}}%
                           {\item[\labelsatz{#1}\hfill]%
                            \settowidth{\hilfslaenge}{\labelsatz{#1}}}%
                         \ifthenelse{\lengthtest{\maximum < \hilfslaenge}}%
                           {\setlength{\maximum}{\hilfslaenge}%
                            \ifthenelse{\equal{#1}{}}%
                               {\renewcommand{\Maximum}{\stdlabel}}%
                               {\renewcommand{\Maximum}{#1}}}%
                           {}%
                      }      
\makeatletter
\newenvironment{AutoLabelLaengenListe}[2][]%
               {\begin{list}%
                      {\labelsatz{#1}\hfill}%
                      {\stepcounter{listennr}%
                       \settowidth{\leftmargin}{M\labelsatz{\ref{listnr\arabic{listennr}}}}%
                       \settowidth{\labelwidth}{\labelsatz{\ref{listnr\arabic{listennr}}}}%
                       \settowidth{\labelsep}{M}%
                       \settowidth{\stdlabellaenge}{\labelsatz{#1}}%
                       \renewcommand{\stdlabel}{#1}%
                       #2%
                       \renewcommand{\Maximum}{}%
                      }%
                }%
               {\renewcommand{\@currentlabel}{\Maximum}%
                \label{listnr\arabic{listennr}}%
                \end{list}%
                }%
\makeatother
\newenvironment{StandardEinrueckung}[2]%
               {\begin{list}%
                      {#1}%
                      {\settowidth{\leftmargin}{M#1}%
                       \settowidth{\labelwidth}{#1}%
                       \settowidth{\labelsep}{M}%
                       #2%
                      }%
                \item}%
               {\end{list}}%
\newenvironment{Einrueckungpur}[1]%
               {\begin{StandardEinrueckung}{#1}{\ListNullAbstaende}}%
               {\end{StandardEinrueckung}}%
\newenvironment{Einrueckung}[1]%
               {\begin{StandardEinrueckung}{#1}{\setlength{\parsep}{0pt}}}%
               {\end{StandardEinrueckung}}%

\makeatletter
\newcommand{\EineNumZeileGleichung}[2][0.5ex]
           {
            
            \vspace{#1} 
            \noindent
            \stepcounter{equation}
            \renewcommand{\@currentlabel}{\arabic{equation}}%
            \phantom{(\arabic{equation})}\hspace*{\fill}
            $\displaystyle{#2}$
            \hspace*{\fill}
            (\arabic{equation})

            \vspace{#1} 
            
           }
\makeatother
\makeatletter
\newcommand{\EineErwNumZeileGleichung}[2][0.5ex]
           {
            
            \vspace{#1} 
            \noindent
            \stepcounter{equation}
            \renewcommand{\@currentlabel}{\arabic{equation}}%
            \phantom{(\arabic{equation})}\hspace*{\fill}
            #2 %
            \hspace*{\fill}
            (\arabic{equation})

            \vspace{#1} 
            
           }
\makeatother
\newcommand{\breitrel}[1]{\hspace*{\tabcolsep} #1 \hspace*{\tabcolsep}}
\newlength{\abstaug}              %
\newenvironment{AllgUnnumGleichung}[2][1.0ex]
               {
  
                \setlength{\abstaug}{#1}
                \vspace{\abstaug}
                \hspace*{\fill}
                $\begin{array}[t]{#2}
                }%
               {\end{array}$
                \hspace*{\fill}
  
                \vspace{\abstaug}

                }%
\newenvironment{AllgNumGleichung}[2][0.0ex]
               {
  
                \setlength{\abstaug}{#1}
                \vspace{\abstaug}
                $\begin{tabular*}{\textwidth}[t]{#2}
                }%
               {\end{tabular*}$

                \vspace{\abstaug}

               }%
\newenvironment{StandardUnnumGleichungKlein}[1][0ex]
               {\renewcommand{\s}{\\[#1] }%
                \begin{AllgUnnumGleichung}{rcl}}%
               {\end{AllgUnnumGleichung}}%
\newcommand{\s}{\\[0ex] }             %
\newenvironment{StandardUnnumGleichung}[1][0ex]%
               {\renewcommand{\s}{\\[#1] }%
                \begin{AllgUnnumGleichung}{>{\displaystyle}rc>{\displaystyle}l}}%
               {\end{AllgUnnumGleichung}}%
\newenvironment{XrelYZNumGleichung}[1][0ex]
               {\renewcommand{\s}{\\[#1] }%
                \begin{AllgNumGleichung}{rcll}}%
               {\end{AllgNumGleichung}}%
\newcommand{\erl}[1]{\hfill\mbox{\hspace*{1.5em}\small (#1)}}

\newcommand{\erllang}[2][0.5\textwidth]%
              {\hfill\hspace*{1.5em}%
               \begin{minipage}[t]{#1}{\small%
                          \begin{list}{(}{\ListNullAbstaende%
                                          \settowidth{\leftmargin}{(}%
                                          \settowidth{\labelwidth}{(}%
                                          \settowidth{\labelsep}{}%
                                         }%
                          \item#2)%
                          \end{list}}%
               \end{minipage}\\[-0.9ex]
              }%
\newcommand{\DefBemUmgeb}[1]%
           {\newenvironment{#1}[1][]%
                           {\begin{Einrueckung}{{\bf #1}}%
                            \ifx##1\empty\else{{\bf ##1}
                            
                                                        }\fi%
                            }%
                           {\end{Einrueckung}}}
\newcommand{\DefSBemUmgeb}[2]
           {\newenvironment{#1}[1][]%
                           {\begin{Einrueckung}{{\bf #2}}%
                            \ifx##1\empty\else{{\bf ##1}
                            
                                                        }\fi%
                            }%
                           {\end{Einrueckung}}}
\makeatletter
\newcommand{\DefBspUmgeb}[3]
           {\newcounter{#2}[#3]%
            \newenvironment{#1}[1][]%
                           {\stepcounter{#2}%
                            \renewcommand{\ZaehlerMarke}{\arabic{#2}}%
                            \renewcommand{\Einzugsname}{{\bf #1 \ZaehlerMarke}}%
                            \begin{Einrueckung}{\Einzugsname}
                            \ifx##1\empty\else{{\bf ##1}\\}\fi%
                            \renewcommand{\@currentlabel}{\ZaehlerMarke}%
                            }%
                           {\end{Einrueckung}}}
\makeatother
\newcommand{\ZaehlerbisEbene}{section}
\newcommand{\Ebenea}{section}
\newcommand{\Ebeneb}{subsection}

\newcommand{\Abschnittnummer}{%
            \ifx\ZaehlerbisEbene\Ebenea{\arabic{section}}%
             \else{%
              \ifx\ZaehlerbisEbene\Ebeneb{\arabic{section}.\arabic{subsection}}%
               \else{\arabic{section}.\arabic{subsection}.\arabic{subsubsection}}%
              \fi}%
            \fi}     
\newcommand{\Abschnittnummerpunkt}{\Abschnittnummer.}     
\newcommand{\Einzugsname}{}
\newcommand{\ZaehlerMarke}{}
\makeatletter
\newcommand{\DefThmUmgeb}[3]%
           {\newcounter{#1}[#3]%
            \newenvironment{#1}[1][]%
                           {\stepcounter{#2}%
                            \setcounter{#1}{\value{#2}}%
                            \renewcommand{\ZaehlerMarke}{\Abschnittnummerpunkt\arabic{#1}}%
                            \renewcommand{\Einzugsname}{{\bf #1 \ZaehlerMarke}}%
                            \begin{Einrueckung}{\Einzugsname}
                            \ifx##1\empty\else{{\bf ##1}
                            
                                                        }\fi%
                            \renewcommand{\@currentlabel}{\ZaehlerMarke}%
                            }%
                           {\end{Einrueckung}}}
\makeatother
\makeatletter
\newcommand{\DefSThmUmgeb}[4]%
           {\newcounter{#1}[#3]%
            \newenvironment{#1}[1][]%
                           {\stepcounter{#2}%
                            \setcounter{#1}{\value{#2}}%
                            \renewcommand{\ZaehlerMarke}{\Abschnittnummerpunkt\arabic{#1}}%
                            \renewcommand{\Einzugsname}{{\bf #4 \ZaehlerMarke}}
                            \begin{Einrueckung}{\Einzugsname}
                            \ifx##1\empty\else{{\bf ##1}

                                                        }\fi%
                            \renewcommand{\@currentlabel}{\ZaehlerMarke}%
                            }%
                           {\end{Einrueckung}}}
\makeatother
\makeatletter
\newcommand{\DefUnterNumThmUmgeb}[5]%
           {\newcounter{#1}[#3]%
            \newcounter{#4}%
            \newenvironment{#1}[1][]%
                           {\ifx##1\empty\else{\stepcounter{#2}\setcounter{#4}{0}}\fi%
                            \stepcounter{#4}%
                            \setcounter{#1}{\value{#2}}%
                            \renewcommand{\ZaehlerMarke}{\Abschnittnummerpunkt\arabic{#1}\alph{#4}}%
                            \renewcommand{\Einzugsname}{{\bf #5 \ZaehlerMarke}}
                            \begin{Einrueckung}{\Einzugsname}
                            \renewcommand{\@currentlabel}{\ZaehlerMarke}%
                            }%
                           {\end{Einrueckung}}}
\makeatother
\newenvironment{Beweis}[1][]%
               {\begin{Einrueckung}{{\bf Beweis}}%
                \ifx#1\empty\else{{\bf #1}

                                            }\fi%
                }%
               {\end{Einrueckung}%
                }%
\newenvironment{Proof}[1][]%
               {\begin{Einrueckung}{{\bf Proof}}%
                \ifx#1\empty\else{{\bf #1}

                                            }\fi%
                }%
               {\end{Einrueckung}%
                }%
               {\begin{Einrueckung}{{\bf \glqq Beweis\grqq}}%
                \ifx#1\empty\else{{\bf #1}
                
                                            }\fi%
                }%
               {\end{Einrueckung}%
                }%
               {\begin{Einrueckung}{{\bf Begr"undung}}%
                \ifx#1\empty\else{{\bf #1}
                
                                            }\fi%
                }%
               {\end{Einrueckung}%
                }%
\newenvironment{Hinrichtung}%
               {\begin{Einrueckungpur}{$\impliz$}}%
               {\end{Einrueckungpur}}%
\newenvironment{Rueckrichtung}%
               {\begin{Einrueckungpur}{$\invimpliz$}}%
               {\end{Einrueckungpur}}%
               {\begin{Einrueckungpur}{\glqq$\teilmenge$\grqq}}%
               {\end{Einrueckungpur}}%
               {\begin{Einrueckungpur}{\glqq$\obermenge$\grqq}}%
               {\end{Einrueckungpur}}%
               {\begin{Einrueckungpur}{"$\teilmenge$"}}%
               {\end{Einrueckungpur}}%
               {\begin{Einrueckungpur}{"$\obermenge$"}}%
               {\end{Einrueckungpur}}%
\newcommand{\qed}{\nopagebreak\hspace*{2em}\hspace*{\fill}{\bf qed}}
\newcommand{\ARabic}{\arabic}
\newcommand{\Nummerntypa}{\arabic}   
\newcommand{\Nummerntypb}{\alph}
\newcommand{\Nummerntypc}{\roman}
\newcommand{\Nummerntypd}{\Alph}

\newcommand{\Nra}{\Nummerntypa{Nummera}}            %
\newcommand{\Nrb}{\Nummerntypb{Nummerb}}            %
\newcommand{\Nrc}{\Nummerntypc{Nummerc}}                
\newcommand{\Nrd}{\Nummerntypd{Nummerd}}                
\newcommand{\ZeichenzuNrTyp}[1]%
           {\ifx#1\ARabic {.}\else{)}%
                  \fi}                              %
\newcommand{\NrZeicha}{\ZeichenzuNrTyp{\Nummerntypa}}
\newcommand{\NrZeichb}{\ZeichenzuNrTyp{\Nummerntypb}}
\newcommand{\NrZeichc}{\ZeichenzuNrTyp{\Nummerntypc}}
\newcommand{\NrZeichd}{\ZeichenzuNrTyp{\Nummerntypd}}
\newcommand{\ListMarkea}%
           {\Nra\NrZeicha}
\newcommand{\ListMarkeb}%
           {\Nra\NrZeicha\Nrb\NrZeichb}
\newcommand{\ListMarkec}%
           {\Nra\NrZeicha\Nrb\NrZeichb\Nrc\NrZeichc}
\newcommand{\ListMarked}%
           {\Nra\NrZeicha\Nrb\NrZeichb\Nrc\NrZeichc\Nrd\NrZeichd}
\newcommand{\Anfangszeichen}{}
\newcommand{\Anfangspunkt}{}
\newcounter{Schachtelebene}
\newcounter{Hilfszaehler}
\newcommand{\Hilfsbefehl}{}
\newcommand{\Schachtelebene}{\alph{Schachtelebene}}
\makeatletter
\newenvironment{AllgNumerierteListe}[2][]
               {\addtocounter{Schachtelebene}{1}%
		\setcounter{Hilfszaehler}{#2}%
                \renewcommand{\Anfangszeichen}%
                             {\renewcommand{\Hilfsbefehl}{\csname Nummerntyp\Schachtelebene \endcsname}%
                              \Hilfsbefehl{Hilfszaehler}}%
                \renewcommand{\Anfangspunkt}%
                             {\csname NrZeich\Schachtelebene \endcsname}%
                \begin{list}%
                      {\stepcounter{Nummer\Schachtelebene}%
                       \csname Nr\Schachtelebene \endcsname
                       \csname NrZeich\Schachtelebene \endcsname
                       }%
                      {\settowidth{\leftmargin}{M\Anfangszeichen\Anfangspunkt}%
                       \settowidth{\labelwidth}{\Anfangszeichen\Anfangspunkt}%
                       \settowidth{\labelsep}{M}%
                       \setlength{\topsep}{0pt}%
                       \setlength{\parskip}{0pt}%
                       \setlength{\partopsep}{0pt}%
                       \setlength{\itemsep}{0pt}%
                       \setlength{\parsep}{0pt}%
                      }%
                \renewcommand{\@currentlabel}{\csname ListMarke\Schachtelebene \endcsname}%
                }%
               {\ifthenelse{\equal{}{}}{\setcounter{Nummer\Schachtelebene}{0}}{}
                \addtocounter{Schachtelebene}{-1}%
                \end{list}}
\makeatother
\newenvironment{NumerierteListe}[1]
               {\begin{AllgNumerierteListe}{#1}}
               {\end{AllgNumerierteListe}}
\newenvironment{WeiterNumerierteListe}[1]
               {\begin{AllgNumerierteListe}[Weiter]{#1}}
               {\end{AllgNumerierteListe}}

\newcommand{\UnnumAnfangszeichen}{}
\newcounter{UnnumSchachtelebene}
\newcommand{\UnnumSchachtelebene}{\alph{UnnumSchachtelebene}}
\makeatletter
\newenvironment{UnnumerierteListe}%
               {\addtocounter{UnnumSchachtelebene}{1}%
                \renewcommand{\UnnumAnfangszeichen}%
                             {\csname UnnumZeich\UnnumSchachtelebene \endcsname}%
                \begin{list}%
                      {\UnnumAnfangszeichen}%
                      {\settowidth{\leftmargin}{M\UnnumAnfangszeichen}%
                       \settowidth{\labelwidth}{\UnnumAnfangszeichen}%
                       \settowidth{\labelsep}{M}%
                       \setlength{\topsep}{0pt}%
                       \setlength{\parskip}{0pt}%
                       \setlength{\partopsep}{0pt}%
                       \setlength{\itemsep}{0pt}%
                       \setlength{\parsep}{0pt}%
                      }%
                }%
               {\addtocounter{UnnumSchachtelebene}{-1}%
                \end{list}}
\makeatother
\newlength{\fktdefhilfslaenge}
\newcommand{\ohnefktdef}[4]
           {\hspace*{\fill}
            $\begin{array}[t]{ccc}%
            #1 & \nach & #2 \\
            #3 & \auf  & #4
            \end{array}$
            \hspace*{\fill}}
\newcommand{\fktdef}[5]
           {\hspace*{\fill}
            $\begin{array}[t]{cccc}%
            #1: & #2 & \nach & #3 \\    
                & #4 & \auf  & #5
            \end{array}$
            \settowidth{\fktdefhilfslaenge}{$#1$:}
            \hspace*{0.6 \fktdefhilfslaenge}  
            \hspace*{\fill}}
\newcommand{\fktdefpur}[5]
           {$\begin{array}[t]{cccc}%
            #1: & #2 & \nach & #3 \\    
                & #4 & \auf  & #5
            \end{array}$}
\newcommand{\fktdefabgesetztpur}[5]
           {
            
            $\begin{array}[t]{cccc}%
            #1: & #2 & \nach & #3 \\    
                & #4 & \auf  & #5
            \end{array}$
            \settowidth{\fktdefhilfslaenge}{$#1$:}
            \hspace*{0.6 \fktdefhilfslaenge}
            
           }
\newcommand{\fktdefabgesetzt}[5]
           {
           
            \hspace*{\fill}
            $\begin{array}[t]{cccc}%
            #1: & #2 & \nach & #3 \\    
                & #4 & \auf  & #5
            \end{array}$
            \settowidth{\fktdefhilfslaenge}{$#1$:}
            \hspace*{0.6 \fktdefhilfslaenge}  
            \hspace*{\fill}
            
            }
\newcommand{\ohnefktdefabgesetzt}[4]
           {      

            \hspace*{\fill}
            $\begin{array}[t]{ccc}%
            #1 & \nach & #2 \\
            #3 & \auf  & #4
            \end{array}$
            \hspace*{\fill}

            }
\newcommand{\doppelohnefktdefabgesetzt}[6]
           {

            \hspace*{\fill}
            $\begin{array}[t]{ccccc}%
            #1 & \nach & #2 & \nach & #3\\
            #4 & \auf  & #5 & \auf  & #6
            \end{array}$
            \hspace*{\fill}

            }
\newcommand{\anhang}%
           {\appendix
            \sectioninh{Anhang}
            \renewcommand{\Abschnittnummer}{%
                  \ifx\ZaehlerbisEbene\Ebenea{\Alph{section}}%
                  \else{%
                        \ifx\ZaehlerbisEbene\Ebeneb{\Alph{section}.\arabic{subsection}}%
                        \else{\Alph{section}.\arabic{subsection}.\arabic{subsubsection}}%
                        \fi}%
                  \fi}%
            \renewcommand{\Abschnittnummerpunkt}{\Abschnittnummer.}     
            }            
\newcommand{\anhangengl}%
           {\appendix
            \sectioninh{Appendix}
            \renewcommand{\Abschnittnummer}{%
                  \ifx\ZaehlerbisEbene\Ebenea{\Alph{section}}%
                  \else{%
                        \ifx\ZaehlerbisEbene\Ebeneb{\Alph{section}.\arabic{subsection}}%
                        \else{\Alph{section}.\arabic{subsection}.\arabic{subsubsection}}%
                        \fi}%
                  \fi}%
            \renewcommand{\Abschnittnummerpunkt}{\Abschnittnummer.}     
            }

\newcounter{wdhlstufe}
\newcommand{\sectioninh}[1]%
           {\section*{#1}%
            \addcontentsline{toc}{section}{#1}}
\newcommand{\bezeichnung}[3]%
           {\begin{Einrueckungpur}{\hbox to 6em{#1}\hbox to 2.4em{\hfill#2}}
            #3
            \end{Einrueckungpur}}

\newcommand{\doppelteinfach}{e}

\newcommand{\ifdoppelt}[1]{\ifthenelse{\equal{\doppelteinfach}{d}}{#1}{}}
\newcommand{\ifeinfach}[1]{\ifthenelse{\equal{\doppelteinfach}{e}}{#1}{}}

\newlength{\querfhilfsl}              %

\newlength{\hll}

\DefThmUmgeb{Theorem}{Theorem}{\ZaehlerbisEbene}
\DefThmUmgeb{Definition}{Definition}{\ZaehlerbisEbene}
\DefThmUmgeb{Satz}{Theorem}{\ZaehlerbisEbene}
\DefThmUmgeb{Proposition}{Theorem}{\ZaehlerbisEbene}
\DefThmUmgeb{Lemma}{Theorem}{\ZaehlerbisEbene}
\DefThmUmgeb{Folgerung}{Theorem}{\ZaehlerbisEbene}
\DefThmUmgeb{Corollary}{Theorem}{\ZaehlerbisEbene}
\DefThmUmgeb{Vorschrift}{Definition}{\ZaehlerbisEbene}
\DefThmUmgeb{Construction}{Definition}{\ZaehlerbisEbene}
\DefSThmUmgeb{FormSatz}{Theorem}{\ZaehlerbisEbene}{\glqq Satz\grqq} 
\DefThmUmgeb{Vermutung}{Theorem}{\ZaehlerbisEbene}
\DefThmUmgeb{Conjecture}{Theorem}{\ZaehlerbisEbene}
\DefThmUmgeb{Konvention}{Definition}{\ZaehlerbisEbene}
\DefThmUmgeb{Feststellung}{Theorem}{\ZaehlerbisEbene}
\DefUnterNumThmUmgeb{DefinitionZusatzNum}{Definition}{\ZaehlerbisEbene}{DefZN}{Definition}
\DefBspUmgeb{Beispiel}{Beispiel}{subsubsection}
\DefBspUmgeb{Example}{Example}{subsubsection}
\DefBspUmgeb{Frage}{Frage}{section}
\DefBspUmgeb{Question}{Question}{section}
\DefBspUmgeb{Aufgabe}{Aufgabe}{section}
\DefBspUmgeb{Ziel}{Ziel}{section}
\DefBemUmgeb{Bemerkung}
\DefBemUmgeb{Remark}
\DefSBemUmgeb{OffeneFrage}{Offene Frage}
\newcommand{\bdf}{\begin{Definition}}
\newcommand{\edf}{\end{Definition}}
\newcommand{\bvorsch}{\begin{Vorschrift}}
\newcommand{\evorsch}{\end{Vorschrift}}
\newcommand{\bconst}{\begin{Construction}}
\newcommand{\econst}{\end{Construction}}
\newcommand{\bthm}{\begin{Theorem}}
\newcommand{\ethm}{\end{Theorem}}
\newcommand{\bsatz}{\begin{Satz}}
\newcommand{\esatz}{\end{Satz}}
\newcommand{\bprop}{\begin{Proposition}}
\newcommand{\eprop}{\end{Proposition}}
\newcommand{\blem}{\begin{Lemma}}
\newcommand{\elem}{\end{Lemma}}
\newcommand{\bfolg}{\begin{Folgerung}}
\newcommand{\efolg}{\end{Folgerung}}
\newcommand{\bcorr}{\begin{Corollary}}
\newcommand{\ecorr}{\end{Corollary}}
\newcommand{\bfest}{\begin{Feststellung}}
\newcommand{\efest}{\end{Feststellung}}
\newcommand{\bbew}{\begin{Beweis}}
\newcommand{\ebew}{\end{Beweis}}
\newcommand{\bpf}{\begin{Proof}}
\newcommand{\epf}{\end{Proof}}
\newcommand{\bwnum}{\begin{WeiterNumerierteListe}}
\newcommand{\ewnum}{\end{WeiterNumerierteListe}}
\newcommand{\bdfzn}{\begin{DefinitionZusatzNum}}
\newcommand{\edfzn}{\end{DefinitionZusatzNum}}
\newcommand{\bbem}{\begin{Bemerkung}}
\newcommand{\ebem}{\end{Bemerkung}}
\newcommand{\brem}{\begin{Remark}}
\newcommand{\erem}{\end{Remark}}
\newcommand{\bnum}{\begin{NumerierteListe}}
\newcommand{\enum}{\end{NumerierteListe}}
\newcommand{\bunum}{\begin{UnnumerierteListe}}
\newcommand{\eunum}{\end{UnnumerierteListe}}
\newcommand{\bbsp}{\begin{Beispiel}}
\newcommand{\ebsp}{\end{Beispiel}}
\newcommand{\bex}{\begin{Example}}
\newcommand{\eex}{\end{Example}}
\newcommand{\bfrag}{\begin{Frage}}
\newcommand{\efrag}{\end{Frage}}
\newcommand{\bquest}{\begin{Question}}
\newcommand{\equest}{\end{Question}}
\newcommand{\baufg}{\begin{Aufgabe}}
\newcommand{\eaufg}{\end{Aufgabe}}
\newcommand{\bof}{\begin{OffeneFrage}}
\newcommand{\eof}{\end{OffeneFrage}}
\newcommand{\bverm}{\begin{Vermutung}}
\newcommand{\everm}{\end{Vermutung}}
\newcommand{\bconj}{\begin{Conjecture}}
\newcommand{\econj}{\end{Conjecture}}
\newcommand{\bkonv}{\begin{Konvention}}
\newcommand{\ekonv}{\end{Konvention}}
\newcommand{\bglklein}{\begin{StandardUnnumGleichungKlein}}
\newcommand{\eglklein}{\end{StandardUnnumGleichungKlein}}
\newcommand{\bgl}{\begin{StandardUnnumGleichung}}
\newcommand{\egl}{\end{StandardUnnumGleichung}}
\newcommand{\bglrtext}{\begin{XrelYZNumGleichung}}
\newcommand{\eglrtext}{\end{XrelYZNumGleichung}}

\newcommand{\berlgl}{\begin{StandardUnnumGleichung}}
\newcommand{\eerlgl}{\end{StandardUnnumGleichung}}
\newcommand{\beinrueck}{\begin{Einrueckungpur}} 
\newcommand{\eeinrueck}{\end{Einrueckungpur}}
\newcommand{\beinflist}{\begin{EinfachListe}} 
\newcommand{\eeinflist}{\end{EinfachListe}}
\newcommand{\beq}{\begin{equation}}
\newcommand{\eeq}{\end{equation}}
\newcommand{\bhin}{\begin{Hinrichtung}}
\newcommand{\ehin}{\end{Hinrichtung}}
\newcommand{\brueck}{\begin{Rueckrichtung}}
\newcommand{\erueck}{\end{Rueckrichtung}}
\newcommand{\bvl}{\begin{AutoLabelLaengenListe}{\ListNullAbstaende}}
\newcommand{\evl}{\end{AutoLabelLaengenListe}}
\newcommand{\df}[1]{{\bf #1}}

\newlength{\adressabstand}
\DefThmUmgeb{Convention}{Definition}{\ZaehlerbisEbene}
\DefThmUmgeb{Notation}{Definition}{\ZaehlerbisEbene}
\makeatletter
\newenvironment{AllgNumerierteListeWeiter}[2][]
               {\addtocounter{Schachtelebene}{1}%
		\setcounter{Hilfszaehler}{#2}%
                \renewcommand{\Anfangszeichen}%
                             {\renewcommand{\Hilfsbefehl}{\csname Nummerntyp\Schachtelebene \endcsname}%
                              \Hilfsbefehl{Hilfszaehler}}%
                \renewcommand{\Anfangspunkt}%
                             {\csname NrZeich\Schachtelebene \endcsname}%
                \begin{list}%
                      {\stepcounter{Nummer\Schachtelebene}%
                       \csname Nr\Schachtelebene \endcsname
                       \csname NrZeich\Schachtelebene \endcsname
                       }%
                      {\settowidth{\leftmargin}{M\Anfangszeichen\Anfangspunkt}%
                       \settowidth{\labelwidth}{\Anfangszeichen\Anfangspunkt}%
                       \settowidth{\labelsep}{M}%
                       \setlength{\topsep}{0pt}%
                       \setlength{\parskip}{0pt}%
                       \setlength{\partopsep}{0pt}%
                       \setlength{\itemsep}{0pt}%
                       \setlength{\parsep}{0pt}%
                      }%
                \renewcommand{\@currentlabel}{\csname ListMarke\Schachtelebene \endcsname}%
                }%
               {\addtocounter{Schachtelebene}{-1}%
                \end{list}}
\makeatother
\renewcommand{\bwnum}{\begin{AllgNumerierteListeWeiter}}
\renewcommand{\ewnum}{\end{AllgNumerierteListeWeiter}}
\newcommand{\bmp}[1]{\begin{minipage}{7.9ex}\hspace{\fill}#1\hspace*{\fill}\end{minipage}}
\newcommand{\yes}{\textbf{$++$}}
\newcommand{\yesnoninj}{\textbf{$+$}}
\renewcommand{\yes}{\mbox{\boldmath$+$}}
\renewcommand{\yesnoninj}{\mbox{\boldmath$\circ$}}
\newcommand{\no}{\mbox{\boldmath$-$}}

\newcommand{\ausnahme}[2][]{\ph{\tiny\ref{#1}}\hspace*{\fill}#2\hspace*{\fill}{\tiny\ref{#1}}}
\renewcommand{\boundfkt}{C_b}
\newcommand{\invzush}{A_0}
\renewcommand{\invzush}{A_{\bullet}}
\renewcommand{\invzush}{A_{\ast}}

\newcommand{\Bigbetrag}[1]%
           {\Bigl|{#1}\Bigr|}            
\newcommand{\Biggbetrag}[1]%
           {\Biggl|{#1}\Biggr|}            

\newcommand{\Bignorm}[2][]{\Bigl\lVert#2\Bigr\rVert_{#1}}
\newcommand{\bignorm}[2][]{\bigl\lVert#2\bigr\rVert_{#1}}

\newcommand{\Bigsupnorm}[1]{\Bignorm[\infty]{#1}}
\newcommand{\bigsupnorm}[1]{\bignorm[\infty]{#1}}

\newcommand{\Bohr}{{\mathrm{Bohr}}}
\newcommand{\rb}{\R_\Bohr}
\newcommand{\voffen}{{V_{\text{open}}}}
\newcommand{\vkomp}{{V_{\text{comp}}}}
\renewcommand{\voffen}{{V}}
\renewcommand{\vkomp}{{K}}
\newcommand{\compl}[1]{\complement#1}

\newcommand{\bconv}{\begin{Convention}}
\newcommand{\econv}{\end{Convention}}
\newcommand{\bnot}{\begin{Notation}}
\newcommand{\enot}{\end{Notation}}

\DeclareMathOperator{\sgn}{sgn}
\DeclareMathOperator{\realteil}{Re}
\DeclareMathOperator{\imaginaerteil}{Im}
\renewcommand{\Im}{\imaginaerteil}
\renewcommand{\re}{\realteil}
\newcommand{\bpm}{\begin{pmatrix}}
\newcommand{\epm}{\end{pmatrix}}
\newcommand{\ph}[1]{\phantom{#1}}
\DeclareMathOperator{\spec}{spec}
\DeclareMathOperator{\vollspan}{span}
\DeclareMathOperator{\dirvsum}{\boxplus}
\newcommand{\bconvlist}{\bvl{}}
\newcommand{\econvlist}{\evl{}}
\newcommand{\citem}[1][]{\iitem[$\bullet$~~ $#1$]$\ldots$~ }
\begin{document}
\title{Loop Quantization and Symmetry: Configuration Spaces}
\author{Christian Fleischhack\thanks{e-mail: 
            {\tt fleischh@math.upb.de}} \\   
        \\
        {\normalsize\em Institut f\"ur Mathematik}\\[\adressabstand]
        {\normalsize\em Universit\"at Paderborn}\\[\adressabstand]
        {\normalsize\em Warburger Stra\ss e 100}\\[\adressabstand]
        {\normalsize\em 33098 Paderborn}\\[\adressabstand]
        {\normalsize\em Germany}
        \\[-25\adressabstand]}     
\date{September 23, 2014}
\maketitle
\newcommand{\iotarestr}{\iota_\stdrestr}
\newcommand{\stdrestr}{\setabb}
\newcommand{\algrestr}[1][\stdrestr]{\alg_{#1}}
\newcommand{\blgrestr}[1][\stdrestr]{\blg_{#1}}
\newcommand{\clg}{{\mathfrak C}}
\newcommand{\dlg}{{\mathfrak D}}
\newcommand{\elg}{{\mathfrak E}}
\newcommand{\urbildalg}{\dlg}
\newcommand{\fktseteins}{{\mathfrak E}}
\newcommand{\fktsetzwei}{{\mathfrak F}}
\newcommand{\ptfktset}{\dlg}
\newcommand{\set}{\mathbf S}
\newcommand{\topset}{\set}   
\newcommand{\elset}{s} 
\renewcommand{\malg}{\spec \alg} 
\newcommand{\Y}{\mathbf{Y}} 
\newcommand{\ely}{\mathbf{y}} 
\newcommand{\restr}[1]{#1_{\stdrestr}}
\newcommand{\elalg}{a} 
\newcommand{\elblg}{b} 
\newcommand{\elclg}{c} 
\newcommand{\eldlg}{d} 
\newcommand{\elelg}{e} 
\newcommand{\elurbildalg}{\eldlg} 
\newcommand{\elfktseteins}{e}
\newcommand{\elfktsetzwei}{f}
\newcommand{\mclg}{\spec \clg} 
\newcommand{\charabb}{\tau}
\newcommand{\charact}{\charakt}
\newcommand{\disjunion}{\sqcup}
\newcommand{\setabb}{\sigma}
\newcommand{\plus}{p}
\newcommand{\cover}{{\cal U}}
\newcommand{\linf}{\ell^\infty}
\newcommand{\gelf}[1][]{\ifthenelse{\equal{#1}{}}{\widetilde}{G_{#1}}}
\newcommand{\gelftrf}{{\sim}}
\newcommand{\coverbasis}{\cover_{\text{Basis}}}
\newcommand{\lin}{{\text{lin}}}

\begin{abstract}
Given two sets $\set_1, \set_2$ and unital $C^\ast$-algebras $\alg_1$, $\alg_2$
of functions thereon, we show that a map $\setabb : \set_1 \nach \set_2$ can
be lifted to a continuous map $\quer\setabb : \malg_1 \nach \malg_2$ iff
$\setabb^\ast \alg_2 := \{\setabb^\ast f \mid f \in \alg_2\} \teilmenge \alg_1$. 
Moreover, $\quer\setabb$ is unique if existing, and injective iff 
$\setabb^\ast \alg_2$ is dense.

Then,
we apply these results to loop quantum gravity and loop quantum cosmology.
Here, the quantum configuration spaces are indeed spectra of certain
$C^\ast$-algebras $\alg_\cosm$ and $\alg_\grav$, respectively, whereas the 
choices for the algebras diverge in the literature. 
We decide now for all usual choices whether the respective cosmological quantum configuration 
space is embedded into the gravitational one. Typically, there is no embedding,
but one can always get an embedding by defining $\alg_\cosm := C^\ast(\setabb^\ast \alg_\grav)$,
where $\setabb$ denotes the embedding between the classical configuration spaces.
Finally, we explicitly determine $C^\ast(\setabb^\ast \alg_\grav)$ in the homogeneous isotropic case
for $\alg_\grav$ generated by the matrix functions of parallel transports along analytic paths.
The cosmological quantum configuration space obtained this way, 
equals the disjoint union of $\R$ and the Bohr compactification of $\R$,
appropriately glued together.

\end{abstract}


\section{Introduction}
Mathematically rigorous 
quantization of physical models has remained a widely unsolved problem.
In particular, theories like gravity, or gauge field theory in general, still wait for
getting quantized to the full extent. One idea to attack this problem is to
simplify these models, e.g., by considering only highly symmetric
situations. Indeed, such reduced models have much
less degrees of freedom, which is why 
they usually can be quantized easier. At the same time,
one hopes that they exhibit some key aspects of the full theory, and this way one expects to
learn more about its quantization. 
This has also been the main motivation
for the invention of loop quantum cosmology. Here,
in the beginning, homogeneous isotropic models
have been quantized along the methods known from loop quantum gravity.
And, indeed, in contrast to the full theory, even the dynamics
has been widely understood. Nevertheless,
one key point that remained open so far
has been the relation between the full and the reduced quantum theory,
so for instance the r\^ole of symmetric states among general ones. 
The main strategy \cite{d86,d83} of how to construct such states consists of 
three basic steps: 
\begin{enumerate}
\item
Embed the reduced configuration space into that of the full theory
and extend that embedding continuously to the quantum configuration
spaces. Typically, the classical configuration spaces are dense subsets
of their respective quantum configuration spaces.
\item
Identify appropriate algebras of separating continuous 
functions on both the full and the reduced
quantum configuration spaces, usually given by cylindrical functions,
and pull then the extended embedding back to get a mapping between these
two algebras. 
\item
Use Gelfand triple constructions, based on the algebras of the previous step and
based on appropriate measures on the configuration spaces, in order
to get states for both theories. Pairing with the mapping of the second step,
one gets a mapping that typically allows to identify states of the reduced theory with
symmetric states among the states of the full theory.
\end{enumerate}
\noindent
Indeed, this outlined strategy formed the basis for the invention of 
loop quantum cosmology some ten years ago. However, it contains 
a very important gap: Denseness is {\em not}\/ sufficient for the existence of 
a continuation of the classical embedding to the quantum regime. Even worse,
a very simple argument \cite{paper28} shows that in the usual loop quantum cosmology framework
such a continuation just does not exist.

In the present article we are going to put all that into a broader context by summarizing
the general circumstances that admit or prevent continuity. Mostly, we will show that 
the first {\em two}\/ steps above are not independent. In fact, 
changing the algebras changes the quantum configuration
spaces as the latter 
are Gelfand-Naimark spectra of the algebras we started with.
Changing these spaces may also turn non-extendibility into
extendibility and vice versa. 
Here, we will now study the following two (related) questions,
first in the general mathematical formulation, then applied to loop quantum gravity:
\begin{itemize}
\item
Under which circumstances does there exist a continuous extension of the 
classical embedding to the quantum regime?
\item
What choices of reduced quantum configuration spaces 
allow for a continuous extension of the 
embedding of the classical configuration spaces?
\end{itemize}
\noindent
Both questions will be answered explicitly for all standard 
conventions used so far in the loop quantum gravity framework.
More precisely, we will see for which types of graphs in the game
(i.e., analytic, straight, etc.)\ and for which selections of cylindrical functions
we have extendibility or non-extendibility. We will determine
in the latter case, how the algebra and, consequently, the configuration
space of the reduced theory has to be modified in order to be
embeddable also in the quantum regime.
Physically, of course, this is just one part of the story. The second part, the 
transition to the phase space relations, will be discussed only in a 
later article.

\leerezeile

\noindent
The paper is organized as follows:

\begin{itemize}
\item
In Section \ref{sect:basics-top-alg} we investigate, when a (not necessarily continuous)
map $\setabb : \set_1 \nach \set_2$ between two sets $\set_1$ and $\set_2$
can be extended to a continuous map $\quer\setabb$ between their compactifications 
$\quer \set_1$ and $\quer \set_2$. 
Here, each $\quer\set_i$ shall be given by $\malg_i$, with each $\alg_i$ being an arbitrary, 
but fixed unital
$C^\ast$-algebra of bounded functions on $\set_i$; this provides us via Gelfand duality
with natural mappings $\iota_i : \set_i \nach \quer\set_i$.
We will show that a continuous extension $\quer\setabb$ of $\setabb$ with 
$\quer \setabb \circ \iota_1 = \iota_2 \circ \setabb$ exists iff 
\bgl
\setabb^\ast \alg_2 \ident \{\setabb^\ast f \mid f \in \alg_2\} & \teilmenge & \alg_1\,.
\egl
This remains true even if $\setabb^\ast \alg_2$ is replaced by $\setabb^\ast \blg_2$
as long as $\blg_2 \teilmenge \alg_2$ generates $\alg_2$ as a $C^\ast$-algebra.
Moreover, the map $\quer\setabb$ is unique if it exists, and it
is injective iff $\setabb^\ast \alg_2$ is dense in $\alg_1$. 
\item
In Section \ref{sect:appl-lqg}, we apply the results of 
Section \ref{sect:basics-top-alg} 
to loop quantum gravity (LQG) and loop quantum cosmology (LQC).
Classically, the configuration spaces are given by the set $\A$ of all 
connections%
\footnote{We may ignore gauge transformations in this paper.} 
in an appropriate principal fibre bundle for the full theory,
and by the set of all symmetric connections in the cosmological case.
For the homogeneous isotropic $k=0$ model, the latter one is just a line in $\A$ 
to be identified with $\R$.
So far, however, many different technical
choices have been made to specify the algebras that define the quantum configuration spaces.
We explicitly identify those combinations that
allow for a continuous embedding of the quantum cosmological configuration space
into that of the full loop quantum gravity theory. It will turn out that most assumptions used so far
lead to non-embedding results.
\item
In Section \ref{sect:new-confsp-lqc}, we outline how the configuration space of 
loop quantum cosmology has to be changed if  one wants to get it
naturally embedded into that of loop quantum gravity. Here, we restrict
ourselves to the most prominent case of the algebra $\alg_\grav$ generated by all the 
parallel transport matrix functions along piecewise analytic loops.
In view of the embeddability criterion from Section \ref{sect:basics-top-alg},
one should {\em define}\/ the algebra $\alg_\cosm$ to be the completion of 
$\setabb^\ast \alg_\grav$.
However, doing this leads to a replacement of the LQC configuration
space. In fact, instead of the so-far standard Bohr compactification $\rb$ of $\R$,
we get the twisted sum of $\R$ and $\rb$ by means of \cite{paper46}.
We prove this somewhat technically by explicitly determining the $C^\ast$-algebra generated
by the parallel transport matrix functions for homogeneous isotropic connections
over $\R^3$. It will turn out to be the $C^\ast$-algebra of almost periodic functions on $\R$ 
plus that of all continuous functions on $\R$ that vanish at infinity.
\end{itemize}

\noindent
Mathematical physicists interested mainly in the applications to loop quantum gravity, 
may check the notations of Section \ref{sect:basics-top-alg} 
as well as Definitions \ref{def:restr_alg} and \ref{def:continuation}
first. Then they may go directly to Section \ref{sect:appl-lqg}. There the most relevant statements
from the preceding section 
(Theorem \ref{thm:general-embedding-criterion-neu} and Proposition \ref{prop:rendall})
can be applied without following their proofs.

\section{Spectral Extension of Mappings}
\label{sect:basics-top-alg}
In this section, we are going to investigate when a map $\setabb : \set_1 \nach \set_2$ 
between two sets can be extended to a continuous
embedding $\quer\setabb : \quer\set_1 \nach \quer\set_2$, where
$\quer\set_1$ and $\quer\set_2$ are certain (locally) compact
spaces that ``(locally) compactify'' $\set_1$ and $\set_2$. More explicitly,
these ``compactifications'' are spectra of certain $C^\ast$-algebras
of functions on $\set_1$ and $\set_2$, respectively. 
For this, we first summarize the relevant properties from topology and $C^\ast$-algebras.
The statements not proven here can, e.g., be found in \cite{Murphy,EMS122} 
in the $C^\ast$-algebraic case,
\cite{Kelley} concerning general topology,
or \cite{BourbakiGenTop1} for uniform structures.

\neueseite

\subsection{Gelfand Transform}

\newcommand{\raum}{X}
\newcommand{\simplefkt}{f}

\bdf
\bunum
\item
For any element $\elalg$ of an abelian $C^\ast$-algebra%
\footnote{We always assume algebras to be nontrivial.} $\alg$,
\bgl
\fktdefabgesetzt{\gelf\elalg}{\malg}{\C}{\charact}{\charact(\elalg)}
\egl
denotes its 
\df{Gelfand transform} $\gelf \elalg$.
\item
The topology on $\malg$ is defined to be the initial topology 
generated by all Gelfand transforms $\gelf\elalg$ with $\elalg \in \alg$.
\eunum
\edf
The celebrated Gelfand-Naimark theorem tells us that 
the Gelfand transform
\bgl
\fktdefabgesetzt{\gelftrf}{\alg}{C_0(\malg)}{\elalg}{\gelf \elalg}
\egl
\noindent
is an isometric $\ast$-isomorphism. 
We usually write $\gelf{\alg}$ for
$\gelftrf(\alg)$.

Moreover, the choice of the initial topology provides us with a useful criterion for
the continuity of functions ranging in $\malg$.

\blem
\label{lem:cont-crit-gelfand}

Let $\raum$ be a topological space and $f : \raum \nach \malg$.
Moreover, let $\blg \teilmenge \alg$ be any subset that generates $\alg$ as
a $C^\ast$-algebra.
Then we have:

\bgl
\text{$\simplefkt$ continuous} 
 & \aequ & 
 \text{$\gelf\elalg \circ f$ continuous for all $\elalg \in \alg$} \\
 & \aequ & \text{$\gelf\elblg \circ f$ continuous for all $\elblg \in \blg$\,.}
\egl
\elem

\bpf
The first equivalence is simply the continuity criterion for 
functions mapping to an initial-topology space.  
To see the second equivalence, observe that the algebra 
operations are continuous, whence we may assume that $\blg$ is
dense in $\alg$.
Now, writing $\elalg \in \alg$ as $\elalg = \lim \elblg_k$ with 
$\elblg_k \in \blg$, we get
\bgl
\supnorm{\gelf\elblg_k \circ \simplefkt - \gelf\elalg \circ \simplefkt}
 \breitrel\leq \supnorm{\gelf\elblg_k - \gelf\elalg}
 \breitrel= \norm{\elblg_k - \elalg}
 \breitrel\gegen 0
\egl
by linearity and isometry of the Gelfand transform. Consequently,
$\gelf\elalg \circ \simplefkt$ is continuous. The reversed implication
is trivial.
\qed
\epf

\subsection{Notations}

\bnot
\label{not:alg-etc}
Unless specified more precisely, 
we let be:
\bconvlist
\citem[\set] a set;
\citem[\linf(\set)] the abelian $C^\ast$-algebra%
\footnote{w.r.t.\ the supremum norm and pointwise addition, multiplication, and inversion.}
of all bounded functions on $\set$;
\citem[\blg] a subset of $\linf(\set)$ such that $\bigcap_{\elblg\in\blg} \elblg^{-1}(0)$ is empty;
\citem[\vollspan \blg] the $\ast$-subalgebra of $\linf(\set)$ generated by $\blg$;
\citem[C^\ast(\blg)] the $C^\ast$-subalgebra of $\linf(\set)$ generated by $\blg$;
\citem[\alg] the $C^\ast$-algebra $C^\ast (\blg)$.
\econvlist
Analogously, $\set_i$, $\blg_i$ and $\alg_i$ 
are defined.

\enot
\noindent
Note that each unital subset $\blg$ of $\linf(\set)$ fulfills the requirement
$\bigcap_{\elblg\in\blg} \elblg^{-1}(0) = \leeremenge$ given above, i.e.,
for all $\elset\in\set$ there is some $\elblg\in\blg$ such that $\elblg(\elset) \neq 0$.

\neueseite


\subsection{Certain Mappings to the Spectrum of a $C^\ast$-Algebra}

\bprop
\label{prop:rendall}
Define%
\footnote{In the following, if necessary, $\iota$ inherits the index from $\alg$.}
the \df{natural mapping} $\iota:\set \nach \malg$ by
\bgl
\fktdefabgesetzt{\iota(\elset)}{\alg}{\C.}{\elalg}{\elalg(\elset)} 
\egl
Then we have:
\bnum{2} 
\item 
$\iota$ is well defined. 
\item 
$\iota(\set)$ is dense in $\malg$. 
\item
\label{item:rendall-same-pts-sep}
$\iota$ separates the same points as $\blg$ does.
\item 
$\iota$ is injective iff $\blg$ separates the points in $\set$. 
\item
\label{item:rendall-continuous}
$\iota$ is continuous iff $\blg$ consists of continuous functions on $\set$ only.
\enum 
Here, for the final assertion, we assumed $\set$ to be given some topology.
\eprop
Before going to prove the proposition, let us state two lemmata.
\blem
\label{lem:gelf-iota=elalg}
For any $\elalg \in \alg$,
we have
\bgl
\gelf \elalg \circ \iota & = & \elalg \,.
\egl
\elem
\bpf\
Observe
$[\gelf \elalg \circ \iota] (\elset) 
 \ident \gelf \elalg (\iota (\elset))
 = [\iota (\elset)] (\elalg)
 = \elalg(\elset)$ for all $\elset \in \set$.
\qed
\epf

\blem
\label{lem:csternvervollst_separation}
For any $\elset,\elset' \in \set$ we have:
\bgl
\elalg (\elset) = \elalg (\elset') \fueralle \elalg \in \alg 
 & \aequ & \elblg (\elset) = \elblg (\elset') \fueralle \elblg \in \blg \,.
\egl
\elem
\bpf
\bhin
Trivial.
\ehin
\brueck
We may assume that $\blg$ is closed w.r.t.\ addition, scalar and algebra multiplication
as well as conjugation.
Now any $\elalg \in \alg$ equals $\lim_{i\gegen\infty} \elblg_i$ for appropriate 
$\elblg_i \in \blg$. Therefore,
$\elalg(\elset) = \lim_{i\gegen\infty} \elblg_i(\elset) = \lim_{i\gegen\infty} \elblg_i(\elset')
 = \elalg(\elset')$.
\qed
\erueck
\epf

\bpf[Proposition \ref{prop:rendall}]
\bnum3
\item
$\iota$ is well defined as $\bigcap_{\elalg\in\alg} \elalg^{-1}(0)$ is empty,
i.e., $\iota(\elset)$ is always nonzero.
\item
\bunum
\item 
Let $\phi : \malg\nach\C$ be continuous with $\phi \ident 0$ on
$\iota(\set)$ and vanishing at infinity. According to the 
Gelfand-Naimark theorem, 
there is an
$\elalg \in \alg$ with $\phi = \gelf \elalg$.  Hence $\phi \ident 0$
from
\bgl
 \elalg \breitrel= \gelf \elalg \circ \iota
        \breitrel= \phi \circ \iota 
        \breitrel= 0.
\egl
\item
Let now $\charact \in \malg\setminus\quer{\iota(\set)}$. 
As $\quer{\iota(\set)}$ is closed, $\malg\setminus\quer{\iota(\set)}$
is a neighbourhood of $\charact$. As, moreover, $\malg$ is 
locally compact Hausdorff, 
there is 
a continuous 
$\phi : \malg\nach\C$ vanishing at infinity and with
$\phi \ident 0$ on $\quer{\iota(\set)}$ and
$\phi(\charact) \neq 0$.
This is impossible as shown above.
\eunum 
\item
For any $\elset_1,\elset_2\in\set$, we have
\bgl
\iota(\elset_1) = \iota(\elset_2) 
 & \aequ & \forall \elalg\in\alg: \elalg(\elset_1) \ident \iota(\elset_1)(\elalg) = \iota(\elset_2)(\elalg) \ident \elalg(\elset_2) \\
 & \aequ & \forall \elblg\in\blg: \elblg(\elset_1) = \elblg(\elset_2) \\
\egl
by Lemma \ref{lem:csternvervollst_separation}.
\item 
For any $\elset_1,\elset_2\in\set$, we have with the preceding step
\bgl
 \elset_1 = \elset_2 
 & \impliz & \forall \elblg\in\blg: \elblg(\elset_1) = \elblg(\elset_2) 
 \breitrel\aequ \iota(\elset_1) = \iota(\elset_2) 
\egl
The first implication is an equivalence for all $\elset_1, \elset_2 \in \set$ 
iff $\blg$ separates the points in $\set$.
This gives the proof.
\item
By Lemma \ref{lem:cont-crit-gelfand}, we see that 
$\iota : \set \nach \malg$ is continuous iff
$\gelf \elblg \circ \iota : \set \nach \C$ is continuous for all $\elblg \in \blg$.
Now the claim follows from Lemma \ref{lem:gelf-iota=elalg}.
\qed
\enum 
\epf
Parts of the proof have been due to \cite{e8}.

\brem
If we had not assumed that the set $\bigcap_{\elblg\in\blg} \elblg^{-1}(0)$ is empty,
$\iota$ could still be defined on the complement $\set_\blg \teilmenge \set$
of this intersection.
The statements prevail after replacing $\set$ by $\set_\blg$.
\erem


\subsection{Uniform Continuity}

Let $\clg$ be an abelian $C^\ast$-algebra. Moreover, we will 
always assume $\C$ to be equipped with the additive uniformity, which is
complete.

\bdf
The \df{standard uniformity} on $\mclg$ is the initial uniformity induced
by all the Gelfand transforms $\gelf \elclg : \mclg \nach \C$.
\edf
In what follows, we always assume $\mclg$ to be given the standard uniformity.
It is compatible with the standard Gelfand-Naimark topology on $\mclg$. 

\blem
\label{lem:unif-crit_initial-top}
Let $\topset$ be a uniform space and let $f : \topset \nach \mclg$ be a mapping.

Then $f$ is uniformly continuous iff $\gelf \elclg \circ f$ is
uniformly continuous for all $\elclg \in \clg$.
\elem
Recall that on a compact Hausdorff space
there is a unique uniformity which is compatible with the topology. 
This uniformity is even complete. 
Moreover, functions from a compact Hausdorff space
to a uniform space are continuous iff they are uniformly continuous. 

\bprop
\label{prop:extcrit_initialtop}
Let $\topset_d$ be a dense subspace of a {\em compact}\/ Hausdorff space $\quer \topset$ 
and let $f$ be a mapping from $\topset_d$ to $\mclg$ with unital $\clg$. Then the following
statements are equivalent:
\bnum2
\item
$f$ can be extended to a continuous function on $\quer \topset$.
\item
$\gelf\elclg \circ f$ can be extended to a continuous function on $\quer\topset$
for all $\elclg \in \clg$.
\enum
\eprop
Note that we have not assumed $f$ to be continuous, a priori. 
\bpf
As $\quer \topset$ is compact and $\mclg$ is complete Hausdorff, 
$f$ can be extended iff $f$ is uniformly
continuous. 
This is equivalent to the uniform continuity of $\gelf\elclg \circ f$ 
by Lemma \ref{lem:unif-crit_initial-top}. This again is equivalent
to the extendibility of $\gelf\elclg \circ f$ for all $\elclg \in \clg$.
\qed
\epf


\subsection{Restriction $C^\ast$-algebras}
\label{subsect:restriction}

\bnot
\label{conv:restriction-allg}
Throughout the remaining section 
we assume to be
\bconvlist
\citem[\setabb]some map from $\set_1$ to $\set_2$;
\citem[\urbildalg]some set of bounded complex-valued functions on $\set_2$.
\econvlist

\enot

\bdf
\label{def:restr_alg}

The \df{restriction} $\setabb^\ast \urbildalg$ of 
the set $\urbildalg$ w.r.t.\ $\setabb$ 
is defined by
\bgl
\setabb^\ast \urbildalg 
 & := & \{\setabb^\ast \elurbildalg \mid \elurbildalg \in \urbildalg\} 
 \breitrel\teilmenge \linf(\set_1) \,.
\egl
\edf
The term ``restriction'' might be misleading in the general case. However,
as we will aim at the case of injective $\setabb$, we opted for that notion.
Obviously, we have

\blem
$\setabb^\ast\urbildalg \teilmenge \linf(\set_1)$ is a $\ast$-subalgebra
if $\urbildalg \teilmenge \linf(\set_2)$ is a $\ast$-subalgebra.
\elem

\blem
$\setabb^\ast (C^\ast(\urbildalg))$ is a dense $\ast$-subalgebra in $C^\ast(\setabb^\ast\urbildalg)$.
\elem
\bpf
\bunum
\item
As $\vollspan$ commutes with the restriction mapping and 
is absorbed by $C^\ast$, we may assume that $\urbildalg$ is a $\ast$-algebra. 
Then, $\setabb^\ast (C^\ast(\urbildalg))$ and $C^\ast(\setabb^\ast\urbildalg)$
are $\ast$-subalgebras of $\linf(\set_1)$.
\item
To show $\setabb^\ast (C^\ast(\urbildalg)) \teilmenge C^\ast(\setabb^\ast\urbildalg)$,
let $\setabb^\ast \elalg_2 \in \setabb^\ast (C^\ast(\urbildalg))$.
By assumption, there are $\elurbildalg_k \in \urbildalg$ with $\elurbildalg_k \gegen \elalg_2$.
This implies $\setabb^\ast\elurbildalg_k \gegen \setabb^\ast\elalg_2$,
hence $\setabb^\ast\elalg_2 \in C^\ast(\setabb^\ast\urbildalg)$.
\item
To show denseness, use 
$\setabb^\ast \urbildalg \teilmenge \setabb^\ast (C^\ast(\urbildalg))$
and the previous item to see
\bgl
C^\ast(\setabb^\ast \urbildalg) 
 & \teilmenge & C^\ast(\setabb^\ast (C^\ast(\urbildalg)))
 \breitrel \teilmenge C^\ast(\setabb^\ast \urbildalg) \,.
\egl
Since $\setabb^\ast (C^\ast(\urbildalg))$ is a $\ast$-subalgebra, we get the proof.
\qed
\eunum
\epf

\bcorr
\label{corr:restr_alg-dicht_gleich_blg-dicht}
We have
\bgl
\text{$\setabb^\ast \alg_2$ is contained in $\alg_1$.}
 & \aequ & \text{$\setabb^\ast \blg_2$ is contained in $\alg_1$.} \\
\text{$\setabb^\ast \alg_2$ is dense in $\alg_1$.}
 & \aequ & \text{$\vollspan \setabb^\ast \blg_2$ is dense in $\alg_1$.} \\
\egl
\ecorr
Recall that $\blg_i$ is some subset of $\linf(\set_i)$
and that $\alg_i$ is the $C^\ast$-subalgebra of $\linf(\set_i)$
generated by $\blg_i$.
\bpf
\bunum
\item
$\setabb^\ast \blg_2 \teilmenge \alg_1$
implies 
$\setabb^\ast \alg_2
   \ident \setabb^\ast (C^\ast(\blg_2))
   \teilmenge C^\ast(\setabb^\ast \blg_2) 
   \teilmenge C^\ast(\alg_1) \ident \alg_1$.
\item
If $\vollspan \setabb^\ast \blg_2$
is even dense in $\alg_1$, we get  
\bgl
\quer{\vollspan \setabb^\ast \blg_2}
   & \teilmenge & \quer{\setabb^\ast \alg_2}
   \breitrel\teilmenge \quer{\alg_1}
   \breitrel\ident \alg_1
   \breitrel= \quer{\vollspan \setabb^\ast \blg_2}\,.
\egl   .
\qed
\eunum
\epf

\bnot
\label{conv:restriction-spez}
Throughout the remaining section 
we assume to be
\bconvlist
\citem[\restr\blg \teilmenge \linf(\set_1)]%
  the restriction $\setabb^\ast \blg_2$ of $\blg_2$;
\citem[\restr\alg \teilmenge \linf(\set_1)]%
  the $C^\ast$-subalgebra of $\linf(\set_1)$ generated by $\setabb^\ast \blg_2$;%
\citem[\restr\iota]%
  the natural mapping  $\restr\iota: \set_1 \nach \spec \restr\alg$.%
\econvlist

\enot

\blem
\label{lem:separation-restriction}
$\restr\iota$ and $\iota_2 \circ \setabb$ separate the same points.%
\footnote{This means
$\restr\iota(\elset_1) = \restr\iota(\elset_1') 
  \aequ [\iota_2 \circ \setabb] (\elset_1) = [\iota_2 \circ \setabb] (\elset_1')$
for any $\elset_1, \elset_1' \in \set_1$.}
\elem
\bpf
For any $\elset_1, \elset_1' \in \set_1$, we have
\bgl
           \restr\iota(\elset_1) = \restr\iota(\elset_1') 
 & \aequ & [\restr\iota(\elset_1)](\restr\elalg) 
           = [\restr\iota(\elset_1')](\restr\elalg) 
                                \fueralle \restr\elalg \in \algrestr[\stdrestr] \\
 & \aequ & \restr\elalg (\elset_1) = \restr\elalg (\elset_1')
                                \fueralle \restr\elalg \in \restr\alg = C^\ast(\restr\blg) \\
 & \aequ & \restr\elblg (\elset_1) = \restr\elblg (\elset_1')
                                \fueralle \restr\elblg \in \restr\blg = \setabb^\ast \blg_2 
				\erl{Lemma \ref{lem:csternvervollst_separation}}\\ 
 & \aequ & \elblg_2 (\setabb(\elset_1)) 
           = \elblg_2 (\setabb(\elset_1')) 
                                \fueralle \elblg_2 \in \blg_2 \\ 
 & \aequ & \elalg_2 (\setabb(\elset_1)) 
           = \elalg_2 (\setabb(\elset_1')) 
                                \fueralle \elalg_2 \in \alg_2 = C^\ast(\blg_2) \\ 
 & \aequ & [\iota_2(\setabb(\elset_1))] (\elalg_2) = [\iota_2(\setabb(\elset_1'))] (\elalg_2)
                                \fueralle \elalg_2 \in \alg_2 \\ 
 & \aequ & [\iota_2 \circ \setabb] (\elset_1) = [\iota_2 \circ \setabb] (\elset_1')
\egl
giving the proof.
\qed
\epf

\blem
Let $\set_1$ be given some topology. Then we have:

\bgl
\text{$\restr\iota$ is continuous.}
 & \aequ & \text{$\iota_2 \circ \setabb$ is continuous.}
\egl
\elem
Note that we do {\em not}\/ require $\setabb$ to be continuous nor
$\set_2$ to carry any topology.
\bpf
Using the observation $\gelf\elalg_2 \circ \iota_2 = \elalg_2$ above, we get
\bgl
\gelf \elalg_2 \circ \iota_2 \circ \setabb
 \breitrel= \elalg_2 \circ \setabb
 \breitrel\ident \setabb^\ast \elalg_2.
\egl
Now, we have
\bgl
 &       & \text{$\restr\iota : \set_1 \nach \restr\malg$ continuous} \\
 & \aequ & \text{$\restr\elblg : \set_1 \nach \C$ continuous $\fueralle \restr\elblg \in \restr\blg$} 
                  \qquad\erl{Proposition \ref{prop:rendall}/\ref{item:rendall-continuous}}\\
 & \aequ & \text{$\setabb^\ast \elblg_2 : \set_1 \nach \C$ continuous $\fueralle \elblg_2 \in \blg_2$} 
                  \erl{since $\restr\blg = \setabb^\ast \blg_2$}\\
 & \aequ & \text{$\gelf \elblg_2 \circ \iota_2 \circ \setabb : \set_1 \nach \C$ continuous $\fueralle \elblg_2 \in \blg_2$} \\
 & \aequ & \text{$\iota_2 \circ \setabb : \set_1 \nach \malg_2$ continuous.}
                  \erl{Lemma \ref{lem:cont-crit-gelfand}}\\
\egl
\qed
\epf


\subsection{Subsets and Supersets of Restriction $C^\ast$-algebras}
\label{subsect:restriction-subsuper}
\newcommand{\moresep}{\succsim}
\newcommand{\smoresep}{\succ}
\newcommand{\lesssep}{\precsim}
\newcommand{\slesssep}{\prec}
\newcommand{\eqsep}{\approx}
\bdf
Let $\fktseteins, \fktsetzwei$ be two sets of 
functions on the same set. 
\bunum
\item
$\fktseteins \lesssep \fktsetzwei$ $\breitrel{:\aequ}$
Any points separated by $\fktseteins$, are also separated by $\fktsetzwei$.
\item
$\fktseteins \eqsep \fktsetzwei$ $\breitrel{:\aequ}$ 
$\fktseteins \lesssep \fktsetzwei$ and $\fktsetzwei \lesssep \fktseteins$.
\eunum
\edf
Analogously, we define the relation for functions.

\blem
\label{lem:teilmenge_impliz_splitting}
With the notations of the preceding definition, we have
\bgl
\fktseteins \teilmenge \fktsetzwei
 & \impliz & \fktseteins \lesssep \fktsetzwei \,.
\egl
\elem
As a reformulation (and slight extension) of Lemma \ref{lem:csternvervollst_separation},
we get
\blem
\label{lem:eqsep_cstern_gen}
We have $\blg \eqsep C^\ast(\blg)$.
\elem

\bcorr
\label{corr:dense-span_iff_eqsep}
Let $\clg$ be an abelian $C^\ast$-algebra containing $\blg$. If $\blg$ contains
$\EINS$, then 
\bgl
\text{$\vollspan \blg$ dense in $\clg$}
 & \aequ & \gelf\blg \eqsep \gelf\clg \,.
\egl
\ecorr
\bpf
First observe that $\vollspan \blg$ is dense in $\clg$ iff
$\vollspan \gelf\blg$ is dense in $\gelf\clg$ by the isomorphy
of the Gelfand transform (considered in both cases w.r.t.\ $\clg$).
\bunum
\item
The assertion for dense $\vollspan \blg$ follows from Lemma \ref{lem:eqsep_cstern_gen} above.
\item
If now $\gelf\blg \eqsep \gelf\clg$, the assertion follows from the Stone-Weierstra\ss\
theorem.
\qed
\eunum
\epf

\bdf
\label{def:continuation}
We define a map 
\bgl
\quer\setabb & : &  \malg_1 \nach \malg_2
\egl
to be an \df{$\alg_1$-continuation} of 
$\setabb : \set_1 \nach \set_2$ iff it fills the diagram
\begin{equation}
\label{eq:cd_setabb-extension-df}
\begin{minipage}[B]{7.8cm}
\vspace*{\CDgap}
\begin{diagram}[labelstyle=\scriptstyle,height=\CDhoehe,l>=3em]
\set_1                 & \relax\rnach^{\setabb}      & \set_2             \\
\relax\dnach_{\iota_1}   &                             & \relax\dnach^{\iota_2} \\
\malg_1                  & \relax\rfillauf^{\quer\setabb} & \malg_2
\end{diagram}
\vspace*{\CDgap}
\end{minipage}\nonumber
\end{equation}
commutatively.
\edf

\blem
\label{lem:extension-formula-1}
If $\quer\setabb$ is 
an $\alg_1$-continuation of $\setabb$, then
we have, for all $\elalg_2 \in \alg_2$,
\bgl
\gelf \elalg_2 \circ \quer\setabb \circ \iota_1
   & = & \setabb^\ast \elalg_2 \,. 
\egl
\elem
\bpf
Use $\gelf\elalg_2 \circ \iota_2 = \elalg_2$ to get 
$\gelf \elalg_2 \circ \quer\setabb \circ \iota_1
 = \gelf \elalg_2 \circ \iota_2 \circ \setabb
 = \elalg_2 \circ \setabb
 \ident \setabb^\ast \elalg_2$.
\qed
\epf

\blem
\label{lem:restriction1}
Define $\dach \setabb : \iota_1(\set_1) \nach \malg_2$ by
$\dach \setabb \circ \iota_1 := \iota_2 \circ \setabb$.
Then, we have
\bgl
\setabb^\ast \alg_2 \lesssep \alg_1
 & \aequ & \text{$\dach\setabb$ is well defined.} \\
 & \aequ & \text{There is an $\alg_1$-continuation $\quer\setabb : \malg_1 \nach \malg_2$.}
\egl
Moreover, 
any such $\quer\setabb$ coincides with $\dach\setabb$ on $\iota_1(\set_1)$.
\elem
\bpf
We have 
\bgl
\!\!\!\!\!\! &       & \setabb^\ast\alg_2 \lesssep \alg_1 \\
\!\!\!\!\!\! & \aequ & \restr\alg \lesssep \alg_1
  \erl{$\restr\alg = C^\ast(\setabb^\ast\alg_2) \eqsep \setabb^\ast\alg_2$ by Lemma \ref{lem:eqsep_cstern_gen}} \\
\!\!\!\!\!\! & \aequ & \restr\iota \lesssep \iota_1 \erl{Proposition \ref{prop:rendall}/\ref{item:rendall-same-pts-sep}}\\
\!\!\!\!\!\! & \aequ & \iota_2 \circ \setabb \lesssep \iota_1 \erl{Lemma \ref{lem:separation-restriction}}\\
\!\!\!\!\!\! & \aequ & \bigl[\text{$\iota_1(\elset_1) = \iota_1(\elset'_1) 
                   \impliz \iota_2(\setabb(\elset_1)) = \iota_2(\setabb(\elset'_1))$}\bigr]%
		     \fueralle \elset_1,\elset'_1 \in \set_1 \\

\!\!\!\!\!\! & \aequ & \text{$\dach\setabb$ well defined on $\iota_1(\set_1)$\,.} \\
\egl
Obviously, the restriction of any $\alg_1$-continuation 
$\quer\setabb$ to $\iota_1(\set_1)$ equals $\dach\setabb$, whence
the second equivalence is trivial.
\qed
\epf

\bcorr
\label{corr:gelfa2_setabb_equals_gelfsetabba2}
Let $\setabb^\ast \alg_2 \teilmenge \alg_1$.
Then we have for any continuous $\alg_1$-continuation $\quer\setabb$
\bgl
\gelf \elalg_2 \circ \quer\setabb
   & = & \gelf{\setabb^\ast \elalg_2} 
   \quad\quad \text{for all $\elalg_2 \in \alg_2$} \s
\egl
and
\bgl
\quer\setabb (\charakt_1) & = & \charakt_1 \circ \setabb^\ast
   \quad\quad \text{for all $\charact_1 \in \malg_1$.}

\egl
\ecorr
\bpf
As $\setabb^\ast \alg_2 \teilmenge \alg_1$,
each $\setabb^\ast \elalg_2$ with $\elalg_2 \in \alg_2$ 
has a well-defined Gelfand transform w.r.t.\ $\malg_1$. By
Lemma \ref{lem:gelf-iota=elalg}
and Lemma \ref{lem:extension-formula-1},
it fulfills
\bgl
 \gelf{\setabb^\ast \elalg_2} \circ \iota_1
 \breitrel= \setabb^\ast \elalg_2
 \breitrel= \gelf \elalg_2 \circ \quer\setabb \circ \iota_1 .
\egl
Hence, $\gelf{\setabb^\ast \elalg_2}$ coincides with $\gelf \elalg_2 \circ \quer\setabb$
on $\iota_1(\set_1)$. As both functions are continuous on $\malg_1$, we get the first assertion
from the denseness of $\iota_1(\set_1)$ in $\malg_1$. The second one follows with
\bgl\
[\quer\setabb(\charact_1)](\elalg_2)
 \breitrel\ident \gelf\elalg_2(\quer\setabb(\charact_1))
 \breitrel= \gelf{\setabb^\ast \elalg_2} (\charact_1)
 \breitrel= \charact_1(\setabb^\ast \elalg_2)
\egl
for all $\elalg_2 \in \alg_2$.
\qed
\epf

\bprop
\label{prop:extension-crit}
The following statements are equivalent, provided $\alg_1$ and $\alg_2$ are unital:
\bnum3
\item
\label{item:restr_einbett}
$\setabb^\ast \alg_2 \teilmenge \alg_1$.
\item
\label{item:ex_cont_ext}
There is a continuous $\alg_1$-continuation of $\setabb$.
\item
\label{item:ex_unique_cont_ext}
There is a unique continuous $\alg_1$-continuation of $\setabb$.
\enum
\eprop

\bpf
\bvl{}
\iitem[\ref{item:restr_einbett} $\impliz$ \ref{item:ex_cont_ext}]
We have $\setabb^\ast \alg_2 \lesssep \alg_1$ by Lemma \ref{lem:teilmenge_impliz_splitting}.
Now, $\dach \setabb \circ \iota_1 := \iota_2 \circ \setabb$
provides us, by Lemma \ref{lem:restriction1}, with a well defined map 
$\dach \setabb : \iota_1(\set_1) \nach \malg_2$.
Moreover, for every $\elalg_2 \in \alg_2$, we know that
\bgl
\gelf \elalg_2 \circ \dach\setabb \circ \iota_1 
 & = & \setabb^\ast \elalg_2
 \breitrel= \gelf{\setabb^\ast \elalg_2} \circ \iota_1 \,,
\egl
where the first equality follows from Lemma \ref{lem:restriction1}
with Lemma \ref{lem:extension-formula-1}, and
the second one from $\setabb^\ast\alg_2 \teilmenge \alg_1$.
Therefore, $\gelf{\setabb^\ast \elalg_2} : \malg_1 \nach \C$ is a 
continuous extension of $\gelf \elalg_2 \circ \dach\setabb : \iota_1(\set_1) \nach \C$.
As $\iota_1(\set_1)$ is dense in $\malg_1$, which is compact by unitality of $\alg_1$,
the assertion follows from
Proposition \ref{prop:extcrit_initialtop}.
\iitem[\ref{item:ex_cont_ext} $\impliz$ \ref{item:ex_unique_cont_ext}]
The restrictions of any two continuous $\alg_1$-continuations 
of $\setabb$ coincide on $\iota_1(\set_1)$ as they equal $\dach\setabb$ there.
As $\iota_1(\set_1)$ is dense in $\malg_1$, they even have to coincide everywhere.
\iitem[\ref{item:ex_unique_cont_ext} $\impliz$ \ref{item:restr_einbett}]
Let $\quer\setabb$ be a continuous $\alg_1$-continuation of $\setabb$ and let $\elalg_2 \in \alg_2$. 
Then $\gelf\elalg_2 \circ \quer\setabb : \malg_1 \nach \C$ is continuous, hence,
by compactness of $\malg_1$,
equals $\gelf\elalg_1$ for some $\elalg_1 \in \alg_1$. 
Now,
we have
\bgl
\setabb^\ast \elalg_2 
 & = & \gelf\elalg_2 \circ \quer\setabb \circ \iota_1
 \breitrel= \gelf\elalg_1 \circ \iota_1
 \breitrel= \elalg_1
 \breitrel\in \alg_1 
\egl
by Lemma \ref{lem:extension-formula-1}.
\qed
\evl
\epf

\blem
\label{lem:inject-crit}
Let $\setabb^\ast \alg_2 \teilmenge \alg_1$ with unital $\alg_2$. Moreover, let 
$\quer\setabb$ be the unique continuous $\alg_1$-continuation of $\setabb$.
Then we have:

\bgl
\text{$\quer\setabb$ injective}
 & \aequ & \text{$\setabb^\ast \alg_2$ dense in $\alg_1$.}
\egl
\elem

\bpf
We have
\bgl
\text{$\quer\setabb$ injective}
 & \aequ & \bigl(\quer\setabb (\charact') = \quer\setabb (\charact'') 
                 \breitrel\impliz \charact' = \charact'' \bigr) \\
 & \aequ & \bigl([\quer\setabb (\charact')](\elalg_2) = [\quer\setabb (\charact'')](\elalg_2) \fueralle \elalg_2 \in \alg_2
                 \breitrel\impliz \charact' = \charact'' \bigr) \\
 & \aequ & \bigl(\gelf{\setabb^\ast \elalg_2}(\charakt') = \gelf{\setabb^\ast \elalg_2}(\charakt'') \fueralle \elalg_2 \in \alg_2
                 \breitrel\impliz \charact' = \charact'' \bigr) \\
 & \aequ & \text{$\gelf{\setabb^\ast \alg_2}$ separates $\malg_1$} \\
 & \aequ & \gelf{\setabb^\ast \alg_2} \eqsep \gelf\alg_1\\
 & \aequ & \text{$\setabb^\ast \alg_2$ is dense in $\alg_1$}
 \erl{$\alg_2$ unital, hence $\alg_1$ unital}
\egl
by Corollaries \ref{corr:dense-span_iff_eqsep} and \ref{corr:gelfa2_setabb_equals_gelfsetabba2}.
\qed
\epf

To summarize the main statements:
\bthm
\label{thm:general-embedding-criterion-neu}
The following statements are equivalent for unital $\alg_1$, $\alg_2$ and $\blg_2$:
\bwnum3
\item
\label{item:alg-dicht}
$\setabb^\ast \alg_2$ is a dense subset of $\alg_1$.
\item
\label{item:blg-dicht}
$\setabb^\ast \blg_2$ spans a dense subset of $\alg_1$.
\item
\label{item:ex-fortsetzung}
$\setabb$ can be extended 
to a continuous embedding 
\bgl
\quer\setabb : \malg_1 \nach \malg_2\,.
\egl
\ewnum
Moreover, any of the conditions above implies:
\bnum2
\item
\label{item:unique}
The embedding $\quer\setabb$ is unique.
\item
\label{item:homeo}
The embedding $\quer\setabb$ is a homeomorphism onto its image.
\enum
\ethm
Recall that
$\setabb : \set_1 \nach \set_2$ is some map between some sets
$\set_1$ and $\set_2$, 
that $\blg_i$ is some subset of $\linf(\set_i)$ 
and that 
$\alg_i$ is the $C^\ast$-algebra generated by $\blg_i$,
for $i = 1,2$. Moreover, note that ``span'' above is understood in
the sense of $\ast$-algebras.
\bpf
\bvl{}
\iitem[\ref{item:alg-dicht} $\aequ$ \ref{item:blg-dicht}]
Corollary \ref{corr:restr_alg-dicht_gleich_blg-dicht}.
\iitem[\ref{item:alg-dicht} $\aequ$ \ref{item:ex-fortsetzung}]
Proposition \ref{prop:extension-crit} and
Lemma \ref{lem:inject-crit}.
\iitem[\ref{item:ex-fortsetzung} \:$\impliz$ \ref{item:unique}]
Proposition \ref{prop:extension-crit}.
\iitem[\ref{item:ex-fortsetzung} \:$\impliz$ \ref{item:homeo}]
Since $\alg_1$ is unital, $\malg_1$ is 
compact, whence $\quer\setabb$ is a homeomorphism onto 
its image in the Hausdorff space $\malg_2$.
\qed
\evl

\epf
Note that 
in the theorem we have not required 
the map $\setabb$ itself to be continuous or injective,
nor even the sets $\set_1$ or $\set_2$ to carry any topology.
To illustrate this for non-injective $\setabb$, let $\set_1$ and $\set_2$ be $S^1$,
and $\setabb(x) := 1$ for all $x \in S^1$.
Taking $\alg_2 := C(S^1)$, we have 
\newcommand{\pkt}{\mathrm{pt}}
\bgl
\setabb^\ast\alg_2  
 & = & \{f \circ \setabb : S^1 \nach \C \mid f \in \alg_2\} 
 \breitrel= \{g : S^1 \nach \C \mid \text{$g$ constant}\}
 \breitrel\iso C(\{\pkt\}) \,.\
\egl
\noindent
Setting $\alg_1 := \setabb^\ast\alg_2$, which is already a $C^\ast$-algebra, we see that $\malg_1 = \{\pkt\}$ and
$\malg_2 = S^1$. At the same time, by construction, $\alg_1$ is a dense
subset of $\setabb^\ast\alg_2$, whence the non-injective map 
$\setabb$ has a unique extension $\quer\setabb$ that is continuous, but also injective.
Indeed, $\quer\setabb$ maps $\pkt$ to $1 \in S^1$.
The reason behind is clear: As the set of constant functions on $S^1$ 
misses to separate any two points in $S^1$, these points are all ``collected'' in a 
single point in the spectrum of $\alg_1$. This way, the spectrum 
shrinks the non-injectivity parts of
$\setabb$ to single points. 

To illustrate the other major case, 
that of non-continuous $\setabb$, keep $\set_1 = \set_2 = S^1$ and 
$\alg_2 = C(S^1)$, but consider now the involution $\setabb : S^1 \nach S^1$ with
\bgl
\setabb(x) & := & \begin{cases}
 		   x & \text{ if $\re x \geq 0$} \\
 		   \quer x & \text{ if $\re x < 0$}
		  \end{cases} \:.
\egl
\noindent
Now $\alg_1 := \setabb^\ast \alg_2$ 
contains lots of non-continuous functions on $S^1$,
provided this has been equipped with the standard topology.
Nevertheless, we may identify $\alg_1$ with $C(\setabb^{-1}(S^1))$
and therefore, $\malg_1 = \setabb^{-1}(S^1)$.
Note that $\setabb^{-1}(S^1)$ and $S^1$ coincide as sets, while
the topologies of $\setabb^{-1}(S^1)$ and $S^1$ are different.
Of course, $\quer\setabb : \malg_1 \nach \malg_2$ is now continuous
(and even a homeomorphism). The non-continuity of $\setabb$ is
encoded in the non-continuity of 
\bgl
\iota_1 : S^1 \ident \set_1 & \nach & \malg_1 = \setabb^{-1}(S^1)\,.
\egl
\noindent
In fact, $\iota_1$ itself as a mapping between sets is the identity,
but as a map between $S^1 = \set_1$ and $\setabb^{-1}(S^1) = \malg_1$ 
it is of course not continuous.
So the non-continuity is already absorbed in the embedding $\iota_1$.
Finally, note that we did not really need any direct information about the 
topologies of $\set_1$ and $\set_2$. Only indirectly, by assuming that 
$\alg_2$ consists of continuous functions on $\set_2$, the topology
came into the game. We may have selected this algebra $\alg_2$ by some other reason, 
so we see that the topology is only relevant on the level of spectra.


\section{Applications to Loop Quantum Gravity}
\label{sect:appl-lqg}
Is the configuration space of loop quantum cosmology (densely) embedded into that
of loop quantum gravity extending the embedding of the classical configuration spaces?
Although this sounds like a definite question, the
answer will very much depend. In fact, there are several technically different versions 
of loop quantum gravity that do give different answers. In its original
form based on piecewise analytic loops, there will be no such embedding
\cite{paper28} -- provided the usual form of loop quantum cosmology is taken.
In this section, we are going to identify the different versions of loop quantum 
gravity/cosmology that lead to embedding or non-embedding results, and to determine 
possible modifications of loop quantum cosmology necessary to 
guarantee the embedding property for the respective technical assumptions
loop quantum gravity is based on.


\subsection{Configuration Spaces -- Classical and Quantum}
The classical configuration space of gravity in the Ashtekar formulation is 
the affine space $\A =: \set_2$ of all smooth connections
in some $SU(2)$-principal
fibre bundle over a three-dimensional manifold $M$. Sometimes, the smoothness 
condition is weakened to some Sobolev condition, but this will rather be irrelevant for
our purposes. The quantum configuration space is indeed given
as the spectrum of some $C^\ast$-algebra $\alg_2$.
This algebra is generated by the set $\blg_2$ of 
parallel transport matrix functions
along all paths in $M$ in a certain smoothness category.%
\footnote{Note that we have chosen $\blg_2$ to be rather minimalistic.
Usually, $\alg_2$ is considered to be the completion of the space
of cylindrical functions or to be generated by at least the so-called
spin-network functions.
Recall that the spin-network functions also include
products of matrix elements of parallel transports represented by means
of irreducible representations labelling the edges of a finite piecewise analytic graph.
Cylindrical functions, on the other hand, are functions that depend continuously (or, sometimes, smoothly)
on the parallel transports along the edges in such a graph.
As, however, the extendibility criterion in Theorem \ref{thm:general-embedding-criterion-neu}
shows that any of these choices lead to the same criterion, we prefer
to use our restrictive choice for $\blg_2$, in view of the calculations 
to be done in Section \ref{sect:new-confsp-lqc}.}
It can easily be
checked that these functions
separate the points in $\A$,
for the most frequently used conventions. \cite{a72,a48,a30}

In cosmology, the configuration space $\set_1$ is formed by symmetric connections
only. In the first form of loop quantum cosmology, symmetric meant 
homogeneous isotropic over $M = \R^3$.\ \cite{d48}
There, the configuration space has just been a line embedded (via $\setabb$) in $\A$.
The quantum configuration space is again given by the spectrum of 
a unital $C^\ast$-algebra, now $\alg_1$ being generated by 
some separating set $\blg_1$
of functions on the classical configuration space. Originally, 
the parallel transport matrix elements along straight edges only, have been used
for $\blg_1$. 
In the homogeneous isotropic case, these are periodic functions on $\set_1 \iso \R$,
such that $\alg_1$ consists of just the almost periodic functions on $\R$
having the Bohr compactification $\rb$ as its spectrum. 
However, it turned out \cite{paper28} that $\rb$ is not embedded into $\Ab$,
at least not as long as it shall extend the classical embedding.
The deeper reason behind this was the observation that the parallel
transport along a circle in the base manifold does not depend almost periodically
on $\R$. In our notation, this just means that $\setabb^\ast \blg_2 \teilmenge \alg_1$
is not given although
being a necessary condition for embeddability of $\rb$ into $\Ab$ 
(see Proposition \ref{prop:extension-crit}).
However, Theorem \ref{thm:general-embedding-criterion-neu}
is a guide to guarantee for 
embeddability. As $C^\ast(\setabb^\ast \blg_2) = \alg_1$ is sufficient
and necessary, we might simply \emph{define} $\alg_1$ to be $C^\ast(\setabb^\ast \blg_2)$. 
Indeed, we will determine $C^\ast(\setabb^\ast \blg_2)$ in the case of homogeneous isotropic 
cosmology in Section \ref{sect:new-confsp-lqc}.


\subsection{Technical Parameters}


\subsubsection{Loop Quantum Gravity}
\label{subsubsect:techn-param-lqg}
Let $P$ be a principal fibre bundle over some at least two-dimensional
manifold $M$ with connected compact structure Lie group $\LG$.
We may assume that $P$ is trivial \cite{paper10}.
Moreover, let $\A$ be the set of all smooth connections in $P$.
We denote the parallel transport%
\footnote{Using some {\em global}\/ trivialization and this way ignoring possible topological obstructions,
we will identify parallel transports with homomorphisms from the
groupoid of all paths (modulo some reasonable equivalence relation)
to the structure group $\LG$.
The trivialization subtleties will be irrelevant for our purposes \cite{paper10}.} 
w.r.t.\ $A$ along a (sufficiently smooth) path $\gamma$ in $M$ by 
$h_A(\gamma)$ or $h_\gamma(A)$.
We will now consider the set $\blg_2 \teilmenge \linf(\A)$ 
of all parallel transport matrix functions $(h_\gamma)^i_j$. 
Here, $\gamma$ runs over the set $\Pf$ of
paths in $M$, 
and $i$ and $j$ over all the matrix indices in some fixed faithful representation
of $\LG$. Note that 
the constant function is in $\blg_2$ as given by the trivial path.
Finally, the spectrum of $\alg_2 := C^\ast(\blg_2)$ is denoted by $\Ab$.

The main technical parameter we will adjust, is the choice of the set $\Pf$ of paths 
under consideration. So far, at least the following assumptions have been used:
\begin{center}
\begin{tabular}{l|l|c}
type & includes all paths that are\ldots & reference \\ \hline
$G_\omega$ & piecewise analytic & \cite{a48} \\
$G_\infty$ & piecewise smooth & \cite{d3,d17,paper3} \\
$G_k$ & piecewise $C^k$ & \cite{paper3} \\
$G_{\mathrm{PL}}$ &  piecewise linear & \cite{e16,d85} \\ 
$G_{\Gamma}$ &  in a fixed arbitrary graph & \cite{d100} \\
$G_{\Gamma,\mathrm{PL}}$ & in a fixed piecewise linear graph & \cite{d101} \\
$G_{\mathrm{B}}$ & in the barycentric subdivision of a linear graph & \cite{st4} \\
\end{tabular}
\end{center}
Note that, in \cite{d100}, the authors did not embed their graphs into
a manifold. Moreover, both for $G_{\Gamma}$ and $G_{\Gamma,\mathrm{PL}}$,
the graph might be infinite.

In the following, we will restrict ourselves to the case of $\LG = SU(2)$ and $M = \R^3$.


\subsubsection{Loop Quantum Cosmology}
Over the last some 10 years, several cosmological models have
been studied in the loop framework. Nevertheless, basically, only homogeneous
models have been investigated non-phe\-nom\-e\-no\-log\-i\-cal\-ly. 
So we will restrict ourselves to that case. 

In the additionally isotropic case, described
by Friedmann-Robertson-Walker models,
there remains a single degree of freedom, that can be encoded in the
derivative $c$ of the scale factor of the universe. There is only an additional 
topological parameter that
labels the three different types of space forms: spherical ($k = 1$),
Euclidean ($k = 0$), hyperbolic ($k = -1$). 
To be specific, for $k = 0$, the configuration space 
is spanned by $c \invzush$, where $c$ runs over $\R$ and $\invzush$ is
a fixed homogeneous and isotropic connection, e.g.,
$\invzush = \tau_1 \dd x + \tau_2 \dd y + \tau_3 \dd z$
where the $\tau_i$ are the Pauli matrices.%
\footnote{Invariant connections for models with other symmetries
have been derived in \cite{d110,d112}. There, also bundle issues are discussed.}
Recall that we have assumed the 
underlying bundle to be trivial, admitting to work in a global trivialization.
Thus, $\set_1 = \R$ with a natural embedding $\setabb : \set_1 = \R \nach \A = \set_2$.
When defining the algebra $\alg_1$, 
one does again not consider these connections themselves,
but their parallel transports along certain edges.
Usually, only straight edges have been taken into account. 
In the Euclidean case, the parallel transports 
for such edges $\gamma$ can be written down explicitly;
they equal
\bgl
h_{c \invzush} (\gamma) & = & \e^{- c \invzush(\dot \gamma) l(\gamma)}\,,
\egl\noindent
where $l(\gamma)$ denotes the length of $\gamma$ determined by the 
Euclidean metric on $\R^3$ and $\gamma$ is parametrized w.r.t.\ to arclength.
But, this choice of paths does not give an embedding of the cosmological quantum
configuration space into that of loop quantum gravity.\ \cite{paper28}

Altogether, there are several options for the paths to be studied:
\begin{center}
\begin{tabular}{l|l|c}
type & includes all paths that are\ldots & reference \\ \hline
$C_{\text{same}}$ & the same as in the LQG model & this paper \\
$C_{\mathrm{PL}}$ & piecewise linear & \cite{d85, d48} \\ 
$C_{\mathrm{fixgeo}}$ & parts of a fixed geodesic & \cite{d48} \\
$C_{\mathrm{min}}$ & one of two incommensurable geodesics & \cite{Thiemann} \\
\end{tabular}
\end{center}
Incommensurability means that the lengths of the two geodesics are $\Q$-independent.
Note that piecewise geodesic is nothing but piecewise linear in the $k = 0$ case.
Moreover, we assume that the trivial path is always included to ensure 
unitality.
Finally, $\blg_1 \teilmenge \linf(\R)$ contains the matrix functions
$c \auf h_{c \invzush} (\gamma)^i_j$ with $\gamma$ running over all
admissible paths. 

\enlargethispage{0.2\baselineskip}
\brem
In the anisotropic case for $k = 0$, one replaces the set of connections $c \invzush$
by that of 
\bgl
A_{\vc} & = & c_1 \tau_1 \dd x + c_2 \tau_2 \dd y + c_3 \tau_3 \dd z
\egl
\noindent
with $\vc = (c_1, c_2, c_3) \in \R^3$. One gets immediately
an embedding $\setabb : \R^3 \nach \A$.
Of course, isotropic connections 
are a special case where all components of $\vc$ coincide.
Consequently, the corresponding $C^\ast$-algebra now consists of functions on $\R^3$.
In principle, the path types in the homogeneous case can be studied again, but the 
last two cases do no longer lead to separating algebras meaning that 
the classical configuration space is no longer embedded into the quantum
one. In the following, however, we will restrict ourselves to the isotropic case.
\erem


\subsection{Constellation matrix}

Theorem \ref{thm:general-embedding-criterion-neu} 
provides us with an explicit criterion whether 
the embedding $\setabb : \set_1 \nach \set_2$ can be extended continuously. 
We only have to check whether $\setabb^\ast \blg_2$ is contained in $\alg_1$ or not,
or even dense therein.
Together with the embedding criterion from Proposition \ref{prop:rendall},
we have
\bprop
\label{prop:embed-matrix}
We have for $k = 0$ in the homogeneous isotropic case:
\begin{center}
\begin{tabular}[t]{l||c|c|c|c}
\phantom{$\R \teilmenge \quer\R$}& \bmp{$C_{\text{same}}$} & \bmp{$C_{\mathrm{PL}}$} & \bmp{$C_{\mathrm{fixgeo}}$} & \bmp{$C_{\mathrm{min}}$} \\  \hline\hline
$G_\omega$               & \yes & \no & \no & \no \\ \hline
$G_\infty$               & \yes & \no & \no & \no \\ \hline
$G_k$                    & \yes & \no & \no & \no \\ \hline
$G_{\mathrm{PL}}$        & \yes & \yes & \yes & \no \\ \hline
$G_{\Gamma}$             & \yes & \ausnahme[except2]\no & \ausnahme[except2]\no & \ausnahme[except2]\no \\ \hline
$G_{\Gamma,\mathrm{PL}}$ & \yes & \ausnahme[except3]\yesnoninj & \ausnahme[except3]\yesnoninj &  \ausnahme[except4]\no\\ \hline  
$G_{\mathrm{B}}$         & \yes & \ausnahme[except3a]\yesnoninj & \ausnahme[except3a]\yesnoninj  & \ausnahme[except5]\no \\ 
\end{tabular}
\end{center}
Here the symbols mean:
\bgl
\yes\ph) & \ldots & \text{continuous injective extension of $\setabb$ to quantum level} \\
\yesnoninj\ph k & \ldots & \text{continuous non-injective extension of $\setabb$ to quantum level} \\
\no\ph) & \ldots & \text{no continuous extension of $\setabb$ to quantum level} \\
\egl
Moreover, the following tables indicate whether the respective
natural mappings injectively map the classical configuration spaces
$\A$ and $\R$ into their quantum counterparts
$\Ab$ and $\quer\R$, respectively:
\begin{center}
\begin{tabular}[t]{l||c}
& $\A \teilmenge \Ab$  \\  \hline\hline
$G_\omega$               & yes \\ \hline
$G_\infty$               & yes \\ \hline
$G_k$                    & yes \\ \hline
$G_{\mathrm{PL}}$        & yes \\ \hline
$G_{\Gamma}$             &     \\ \hline
$G_{\Gamma,\mathrm{PL}}$ &     \\ \hline  
$G_{\mathrm{B}}$         &     \\ 
\end{tabular}
\qquad\qquad
\begin{tabular}[t]{l||c}
& $\R \teilmenge \quer\R$ \\ \hline\hline
$C_{\text{same}}$ & \ausnahme[except1]{yes}  \\ \hline
$C_{\mathrm{PL}}$ & yes \\  \hline
$C_{\mathrm{fixgeo}}$ & yes \\ \hline
$C_{\mathrm{min}}$ & yes \\
\end{tabular}
\end{center}
An empty slot means that there is no general answer.
Note that $\Ab$ and $\quer\R$ explicitly depend on the chosen category of paths, i.e.\ the
respective $G$- and $C$-types.

The small numbers in the tables above denote the following exceptions:
\bnum3
\item
\label{except2}
True at least if parallel transports along non-straight paths never depend almost periodically
on $c$. We expect this to be the case, however, do not have a proof for it.
Nevertheless, by \cite{paper28}, generically 
the parallel transport along a non-straight path is not almost periodic; more precisely, 
there is always an initial path such that the parallel transport along any
nontrivial subpath of it is not almost periodic. 
\item
\label{except3}
Injectivity is given if the edge lengths in $\GR$ span $\R$ over $\Z$.
This requires at least a graph with uncountably many edges.
\item
\label{except4}
``\yes'' (or ``\yesnoninj'', resp.) 
iff the edge lengths appearing in $\GR$ have the same (or smaller, resp.) $\Z$-span as those of
the two lengths used for $C_{\mathrm{min}}$.
\item
\label{except3a}
Injectivity is given as in Exception \ref{except3}.
Note that this means that already the starting graph has to be uncountable.
\item
\label{except5}
``\yesnoninj'' iff the graph the subdivision started with, contained a single edge
having a length in the $\Z$-span of the two edge lengths used for $C_{\mathrm{min}}$.
\item
\label{except1}
``no'' in $G_{\GR,\mathrm{PL}}$ and 
$G_{\mathrm{B}}$, respectively, iff all lengths of edges appearing in $\GR$ are commensurable.
Unknown for $G_{\Gamma}$, in general.
\enum

\eprop
\noindent

\brem
\bnum3
\item
Note that the entries for the case $G_{\Gamma}$ are 
given under the assumption that $\Gamma$ is {\em not}\/ piecewise linear.
\item
In the cases where only paths in a fixed (possibly infinite) 
graph $\GR$ are studied at the level $\A$ (i.e., the last three cases),
a general statement on the injectivity is not possible. Nevertheless, a few special
cases can be decided.
If the graph does not form a dense subset of $M$ (e.g., if $\GR$ is finite), 
then $\iota_2$ is not injective 
as the parallel transports along the edges in $\GR$ do not separate 
the points in $\A$. (Consider, e.g., two different smooth connections whose difference
is supported outside $\GR$.)
On the other hand, if the graph is constructed by barycentric subdivision 
of a starting graph and this starting graph is ``sufficiently large'', then 
we have injectivity of $\iota_2$ by the separation property.
\item
Roughly speaking, an ``$\yesnoninj$'' entry means that there are not enough paths used in the 
full theory. It is rather unrealistic that such a combination gives a
reasonable physical theory. Nevertheless, e.g., for the spectral triple 
construction in loop quantum gravity \cite{st4} one has to restrict oneself to
a piecewise linear fixed graph. To investigate possible extensions of this framework 
to cosmology, one should therefore take the same sets of graphs for the reduced and the full 
theory.

On the other hand, a ``$\no$'' entry means that there are not enough paths in the game at the
cosmological level. This can be avoided taking again the same set of paths for both theories
or possibly go over to the piecewise linear theory. We will study the implications for the
former choice more in detail in Section \ref{sect:new-confsp-lqc}.
\enum
\erem

For completeness we include the following lemma that will be needed in the proof
of the proposition above.

\newcommand{\laenge}{l}
\newcommand{\Laenge}{L}

\blem
\label{lem:characts_determine_algebra}
Define
\fktdefabgesetzt{\charact_\laenge}{\R}{\C}c{\e^{\I c \laenge}}
for $\laenge \in \R$ and 
\bgl
\clg(\Laenge) & := & C^\ast(\{\charact_\laenge \mid \laenge \in \Laenge\}) 
\breitrel\teilmenge C_\AP(\R)
\breitrel\teilmenge \boundfkt(\R)
\egl
for any $\Laenge \teilmenge \R$.
Then we have 
\bgl
\laenge \in \vollspan_\Z \Laenge & \aequ & \charact_\laenge \in \clg(\Laenge) \,.
\egl
\elem
Here, $C_\AP(\R)$ denotes the $C^\ast$-algebra of almost periodic functions on $\R$.
\bpf
\bhin
Follows from $\charact_{\laenge_1} \charact_{\laenge_2} = \charact_{\laenge_1 + \laenge_2}$ 
and $\charact^\ast_\laenge = \charact_{-\laenge}$,
implying $\clg(\Laenge) = \clg(\vollspan_\Z \Laenge)$.
\ehin
\brueck
We may assume that $\Laenge$ is closed w.r.t.\ $\vollspan_\Z$.
Suppose $\laenge' \notin \Laenge$. 
If $\charact_{\laenge'}$ was in the unital $\ast$-subalgebra $\dlg$ generated by 
$\{\charact_\laenge \mid \laenge \in \Laenge\}$, then
$\charact_{\laenge'} = \sum_i \alpha_i \charact_{\laenge_i}$ with appropriate $\alpha_i \in \C$ and
$\laenge_i \in \Laenge$. Consequently,
\bglklein
1 \breitrel= \skalprod{\gelf\charact_{\laenge'}}{\gelf\charact_{\laenge'}}
 & = & \sum_i \alpha_i \skalprod{\gelf\charact_{\laenge'}}{\gelf\charact_{\laenge_i}}
 \breitrel= 0 \,,
\eglklein
using that different characters are orthogonal in $L_2(\rb,\mu_\Bohr)$.
Here, Gelfand duality is understood w.r.t.\ $C_\AP(\R)$ having $\rb$ as its spectrum.
If now $\charact_{\laenge'}$ was in $\clg(\Laenge)$, then it can approximated by elements in $\dlg$
in supnorm, hence in $L_2$-norm as well. The contradiction is now obvious.
\qed
\erueck
\epf

\bcorr
With the notations of Lemma \ref{lem:characts_determine_algebra} we have for
$\Laenge_1, \Laenge_2 \teilmenge \R$
\bgl
\text{$\clg(\Laenge_1) \teilmenge \clg(\Laenge_2)$}
 & \aequ & \vollspan_\Z \Laenge_1 \teilmenge \vollspan_\Z \Laenge_2
\egl
and 
\bgl
\text{$\clg(\Laenge_1) \teilmenge \clg(\Laenge_2)$ dense}
 & \aequ & \clg(\Laenge_1) = \clg(\Laenge_2) \\
 & \aequ & \vollspan_\Z \Laenge_1 = \vollspan_\Z \Laenge_2\,.
\egl
\ecorr

\bcorr
\label{corr:L-determines-alg1}
Assume that the sets $\Pf_\cosm$ and $\Pf_\grav$ of paths used in the cosmological and the gravity case,
respectively, consist of linear paths and their concatenations only.
Denote by $\Laenge_\cosm$ and $\Laenge_\grav$ the set of all lengths occurring in $\Pf_\cosm$ and $\Pf_\grav$.
Then 
\bgl
\clg(\Laenge_\cosm) = \alg_1
& \text{ and } & 
\clg(\Laenge_\grav) = \setabb^\ast \alg_2 \,.
\egl
In particular, we have
\bgl
\setabb^\ast \alg_2 \teilmenge \alg_1
 & \aequ & \vollspan_\Z L_\grav \teilmenge \vollspan_\Z L_\cosm
\egl
and 
\bgl
\text{$\setabb^\ast \alg_2 \teilmenge \alg_1$ dense}
 & \aequ & \vollspan_\Z \Laenge_\cosm = \vollspan_\Z \Laenge_\grav \,.
\egl
\ecorr
\bpf
The parallel transport
along a straight line $\gamma$ with $\norm{\dot\gamma} = 1$ 
for the connection $c \invzush$ is given 
by $\e^{-c \invzush(\dot\gamma) \laenge(\gamma)}$.
As $\sin(c \laenge(\gamma))$ and $\cos(c \laenge(\gamma))$
are linear combinations of the matrix elements of that function,
we have $\charact_{\laenge(\gamma)}\in\alg_1$. Hence $\clg(\Laenge) \teilmenge \alg_1$.
On the other hand, such parallel transport functions along straight paths 
generate here the parallel transport functions along arbitrary paths. As 
the former ones are contained in $\clg(L)$ and the latter ones generate 
$\alg_1$, we have $\alg_1 = \clg(L)$.

The case of $\alg_2$, i.e., that of full gravity is completely analogous.
\qed
\epf

\bpf[Proposition \ref{prop:embed-matrix}]
\bunum
\item
To prove the injectivity of $\iota_1 : \R \nach \quer\R$, observe
that in each case (up to Exception \ref{except1} above) 
there exist straight paths of incommensurable lengths. As
they separate the points in $\R$, Proposition \ref{prop:rendall} gives injectivity.
\item
The injectivity of $\iota_2 : \A \nach \Ab$ in the indicated cases 
is proven similarly. Observe here that
the smooth connections in each case are separated by the parallel transport matrix functions
along respectively admitted paths. 
(See Appendix \ref{app:separate} for a proof).

\item
The case $C_{\mathrm{fixgeo}}$
of parts of a fixed geodesic (i.e., parts of a fixed straight line) can be reduced to 
the case $C_{\mathrm{PL}}$ of all piecewise linear paths. 
As the length of partial geodesics runs over all%
\footnote{Obviously, it would even suffice to take 
all paths that are parts of a fixed geodesic of finite positive length.} 
positive numbers, they span full $\R$ w.r.t.\ $\Z$; the same is true for all piecewise linear
paths. Therefore, the algebras $\alg_1$ 
for $C_{\mathrm{fixgeo}}$ and $C_{\mathrm{PL}}$ coincide 
by Corollary \ref{corr:L-determines-alg1},
whence the columns for $C_{\mathrm{fixgeo}}$ and for $C_{\mathrm{PL}}$ are identical.
(In the notation of Section \ref{sect:basics-top-alg}, however, the algebras $\blg_1$ 
do not coincide.) 
\item
The cases with $C_{\text{same}}$ are obvious. In fact, as $\blg_1$ consists
just of the restrictions of all the functions $f \in \alg_2$ 
to $\set_1$, we have $C^\ast(\setabb^\ast\alg_2) = C^\ast(\setabb^\ast\blg_2) = \alg_1$ by construction.
\item
The cases with $G_\omega$, $G_\infty$, $G_k$, but not $C_{\text{same}}$
can be reduced to that studied in \cite{paper28}.
The easiest case is that of a circle $\gamma$ in $\R^2 \teilmenge \R^3$
which, of course, is not a path comprised by the $C$-choices.
Let us assume $\gamma(t) = (\cos t,\sin t,0)$ with $t \in [0,2\pi]$.
A straightforward calculation shows
that 

\begin{eqnarray}
h_{c \invzush} (\gamma)^1_2
 & = & \I^{\frac32} \: \frac{\sin \Bigl(2\pi c \: \sqrt{1 + \inv{4c^2}}\Bigr)}{\sqrt{1 + \inv{4c^2}}}\,.
\nonumber
\end{eqnarray}
(Recall that the indices $1$ and $2$ indicate the respective $SU(2)$ matrix component.)
Obviously, this matrix function is not almost periodic, hence
its restriction to $\set_1 = \R$ is not contained in $\alg_1$.

\item
The case $G_{\mathrm{PL}}$--$C_{\mathrm{PL}}$
coincides with $G_{\mathrm{PL}}$--$C_{\text{same}}$.
\item
The case $G_{\mathrm{PL}}$ and 
$C_{\mathrm{min}}$, however, gives $\setabb^\ast \alg_2 \not\teilmenge \alg_1$.
In fact, the latter one is generated by the functions 
on $\R$ having two incommensurable periods (or being constant). But, 
by Corollary \ref{corr:L-determines-alg1}, this
algebra does not comprise the algebra of all almost-periodic functions 
being $\setabb^\ast \alg_2$. 

\item
\label{case:GR-PL}
The cases with $G_\Gamma$, except for $C_{\text{same}}$, seem to be similar to that
of $G_\omega$. (Recall, that here $\Gamma$ is not piecewise linear.) 
However, the argumentation is much more involved
as so far it is unknown whether parallel transports along non-straight
edges always depend non-almost periodically on $c$ (see Exception \ref{except2}).
Nevertheless, given that conjecture to be true, the statement follows as for
$G_\omega$.

\item
\label{case:GRPL-PL}
In the case $G_{\mathrm{\GR,PL}}$--$C_{\mathrm{PL}}$, 
apply Corollary \ref{corr:L-determines-alg1}:
As $\Laenge_\cosm$ spans $\R$ over $\Z$, we always have 
$\Laenge_\grav \teilmenge \Laenge_\cosm$, hence extendibility.
However, injectivity is given iff the $\Z$-span of the edge lengths in $\GR$ 
is full $\R$.
\item
The case $G_{\mathrm{\GR,PL}}$--$C_{\mathrm{min}}$ is a little bit different.
Unless each edge length appearing in $\GR$ lies in the $\Z$-span of
the two lengths used for $C_{\mathrm{min}}$, there will be 
paths whose parallel transports have the ``wrong'' period in $c$, whence
$\setabb^\ast \alg_2 \not\teilmenge \alg_1$.
\item
The case $G_{\mathrm{B}}$--$C_{\mathrm{PL}}$ 
is similar to 
$G_{\mathrm{\GR,PL}}$--$C_{\mathrm{PL}}$.

\item
For the case $G_{\mathrm{B}}$ and $C_{\mathrm{min}}$,
observe that $\inf \Laenge_\grav$ is zero.
Hence, $\vollspan_\Z \Laenge_\grav$ cannot be contained in $\vollspan_\Z \Laenge_\cosm$.
\qed
\eunum
\epf


\section{New Configuration Space for Loop Quantum Cosmology}
\label{sect:new-confsp-lqc}
In the introduction, we sketched the Bojowald-Kastrup scheme that leads, in principle,
to the quantization of a reduced theory along the 
lines of the full theory. In order to give us a chance to implement this strategy
successfully, the corresponding algebras $\alg_1$ and $\alg_2$
have to fulfill the compatibility condition that $\setabb^\ast \alg_2$ is
a dense subalgebra of $\alg_1$. In the standard LQC-LQG setting, however,
this condition is not met. At the same time, we have seen that simply
replacing $\alg_1$ by (the $C^\ast$-algebra generated by) $\setabb^\ast \alg_2$
solves this problem. In other words, we should just take the same sets of
paths underlying the parallel transports in loop quantum gravity and in loop quantum
cosmology. This, however, will lead to a different configuration space
for loop quantum cosmology. In this section, we are going to determine
this space for the easiest case of homogeneous isotropic cosmology and assume
that the full gravity theory is based on piecewise analytic paths.
For this, we will prove that the parallel transport for homogeneous
isotropic connections $c \invzush$ along any analytic
path $\gamma$
is a sum of a unique%
\footnote{See Corollary \ref{corr:AP+C_0_ist_C*-alg} in the appendix for a proof that 
the almost periodic functions and the continuous functions vanishing at infinity
have trivial intersection.}
continuous function periodic in $c$ and a
unique continuous function vanishing at infinity. Even more,
any such sum is in the $C^\ast$-algebra generated by the parallel transport matrix functions.
In other words,
$\alg_1 = C^\ast(\setabb^\ast\alg_2)$ 
equals $C_0(\R) \dirvsum C_\AP(\R)$.
Now, by \cite{paper46},
its spectrum
is the $\iota$-twisted sum
\bgl
\R \disjunion \rb \,,
\egl
\noindent
a topology on the disjoint union of $\R$ and $\rb$ that intertwines both
spaces in a topologically nontrivial way.


\subsection{Selection of Paths}
\label{subsect:selection-of-paths}
\newcommand{\qm}{m}
\newcommand{\qqm}{{\quer m}}
\newcommand{\vorz}{-}
\newcommand{\invvorz}{+}

As we have already mentioned above, $C^\ast(\setabb^\ast \alg_2)$
equals $C^\ast(\setabb^\ast \blg_2)$ for any 
set $\blg_2$ generating $\alg_2$. This means, a clever (in particular, small)
choice for $\blg_2$ will very much reduce the computational costs 
we will have to pay in the following.

We have already mentioned that the original full LQG algebra $\alg_2 \teilmenge \linf(\A)$
is generated by
all parallel transport matrix functions
\bgl
A & \auf & h_A(\gamma)^i_j\,
\egl\noindent
where $i,j$ are $1$ or $2$ (remember that $\LG = SU(2)$) and $\gamma$ runs, by assumption, over
all piecewise analytic paths in $M$. As parallel transports are homomorphisms on the
path groupoid $\Pf$ and as each piecewise analytic path is a finite product of analytic paths,
$\alg_2$ is already generated by the set $\blg_2$ of all the matrix functions above where $\gamma$ runs
over just the analytic paths in $M$ only. W.l.o.g., we require $\gamma$ to be parametrized w.r.t.\
arclength.

We may shrink this class of paths even further, 
changing $\blg_2$ though, but not $C^\ast(\setabb^\ast \blg_2)$.
Indeed, note that we have
\bgl
\fktdefabgesetzt\setabb\R\A c{c \invzush}
\egl\noindent
with $c \invzush$ running over the homogeneous isotropic 
connections.
Thus, of course, the elements in $\setabb^\ast \blg_2$ only ``see''
the behaviour of the parallel transports on these 
connections. 
If we now, say, rotate a path by a constant matrix, the parallel transport
will change only by some conjugation with a fixed element in $SU(2)$.
If we translate the path, the parallel transport will even remain unchanged.
Therefore, we might restrict ourselves even to a single representative 
from each orbit of the Euclidean group acting on the paths, without changing the algebra.
Rotating and, if necessary, again 
decomposing the paths, we now see that $\setabb^\ast \alg_2$ is generated
by all parallel transport matrix functions
\bgl
c & \auf & h_{c \invzush}(\gamma)^i_j\,,
\egl\noindent
where $\gamma$ runs over all analytic paths in $\R^3$ 
for which $\dot\gamma$ is parallel to the $z$-axis
(unless $\gamma$ is trivial).

To sum up, we agree on
\bnot
\label{conv:dlg}
We denote by $\ptfktset$ the
set of matrix element functions of parallel transports along 
the analytic paths $\gamma$ in $\R^3$ that are parametrized by arclength
and that are not parallel to the $z$-axis (unless trivial).
\enot
Although $\ptfktset$ need not%
\footnote{That is the reason why, in order 
to avoid confusion with Notation \ref{not:alg-etc},
we write $\ptfktset$ instead of $\blg_2$.}
generate
full $\alg_2$,
we have shown
that $C^\ast(\setabb^\ast \ptfktset)$ equals $C^\ast(\setabb^\ast\alg_2)$
which is the algebra we need.


\subsection{Parallel Transport Differential Equation}
\label{subsect:parallel-transport_dgl}
Now, let us derive 
the differential equation \cite{paper28} that gives us the matrix elements of the 
parallel transports for $c \invzush$ along $\gamma$.
We denote the parallel transport along $\gamma$ from $0$ to $t$ w.r.t.\ $c \invzush$
by $g_c(t) \in SU(2)$. The differential equation determining $g_c$ is
\bgl
\dot g_c = - c \invzush (\dot\gamma) \: g_c 
 & \text{ \: with \: } &  g_c(0) = \EINS.
\egl\noindent
Again, we assume
$\invzush = \tau_1 \dd x + \tau_2 \dd y + \tau_3 \dd z$
with Pauli matrices $\tau_i$,
and define $a_c, b_c$ by 
\bgl
g_c
 & =: & \bpm
       \ph+ a_c & b_c \\ -\quer b_c & \quer a_c 
       \epm \,.
\egl\noindent
If confusion is unlikely, the will drop the index $c$.
We will write any path 
$\gamma : I \nach \R^3$ as a coordinate triple $(x,y,z)$ and define
\bgl
   m & := & \dot x \vorz \I \dot y \\
   n & := & \dot z \,.
\egl\noindent
Here, $I \teilmenge \R$ is some interval containing $0$. As $\gamma$
is assumed to be parametrized w.r.t.\ the arclength,
we have
\bgl
{\betrag \qm}^2 + n^2 \breitrel= \norm{\dot\gamma}^2 \breitrel\ident 1 
\egl\noindent
and get after a straightforward calculation \cite{paper28}
\begin{eqnarray}
\label{eq:dgl_apunkt}
\dot a & = & \I c (na - \qm \quer b) \nonumber\\
\dot b & = & \I c (nb + \qm \quer a) \nonumber
\end{eqnarray}
\noindent
with the initial conditions
\begin{eqnarray}
a(0) & = & 1 \nonumber \\
b(0) & = & 0. \nonumber
\end{eqnarray}\noindent
As we consider only paths that are not parallel to the $z$-axis, we
have $m \neq 0$. Therefore,
\begin{eqnarray}
\ddot a & = & \I c (\dot n - Mn) a  - c^2 a + M \dot a
\label{eq:dgl2_a} \\
\ddot b 
  & = & \I c (\dot n - M n) b
         - c^2 b + M \dot b \,. \nonumber
\label{eq:dgl2_b}
\end{eqnarray}
\noindent
with
\bgl
M & := & \frac{\dot \qm}{\qm} \,.
\egl\noindent
\noindent
The first derivative can be removed by factorizing
$a = \sqrt{\qm} \: \faf$ and $b = \sqrt{\qm} \: \fbf$.
This leads to 
\begin{eqnarray}
\label{eq:dgl-faf}
\ddot\faf + c^2 \faf
  & = & \Bigl(\frac14 M^2 - \inv2 \dot M 
                + \I c (\dot n - Mn) \Bigr) \: \faf \\
\label{eq:dgl-fbf}
\ddot\fbf + c^2 \fbf
  & = & \Bigl(\frac14 M^2 - \inv2 \dot M 
                + \I c (\dot n - Mn) \Bigr) \: \fbf 
\end{eqnarray}
and the initial conditions
\begin{eqnarray}
\label{eq:dgl-faf-anfbed0}
    \faf(0) & = & \inv{\sqrt{\qm(0)}} \\
\label{eq:dgl-faf-anfbed1}
\dot\faf(0) & = & \frac{\I c n(0) - \einhalb M(0)}{\sqrt{\qm(0)}} \\
\label{eq:dgl-fbf-anfbed0}
    \fbf(0) & = & 0 \\ 
\label{eq:dgl-fbf-anfbed1}
\dot\fbf(0) & = & \I c \: \sqrt{\qm(0)} \,.
\end{eqnarray}

\noindent
As $a(t)$ and $b(t)$ at given $t$ are up to a nonzero factor equal to $\faf(t)$ and $\fbf(t)$, respectively,
just all these functions above span the same space as $\setabb^\ast\ptfktset$.

Heuristically, the solution for large $\betrag c$ should be periodic in $c$. In fact,
we may consider the coefficient at the right hand side of differential equation 
\eqref{eq:dgl-faf} as a perturbation
of the $c^2$-term at the left hand side,
as the former one grows at most with $\betrag c$. So, the solution should be something periodic plus
something vanishing at infinity. A more careful analysis below will show that this
is basically correct. 

Let us now prove our main results on the spectrum of $\alg_1 = C^\ast(\setabb^\ast \ptfktset)$ 
in two steps.
\begin{enumerate}
\item
Show that there are paths, for which the corresponding solutions 
\bgl
c \auf \faf_c(t)
 & \text{ \: and \: } & 
c \auf \fbf_c(t)
\egl
form a dense subset of $C_\AP(\R) + C_0(\R)$,
where $C_\AP(\R)$ denotes the set of almost periodic functions on $\R$ and
$C_0(\R)$ the set of continuous functions on $\R$ vanishing at $\infty$.
For this, we will need straight lines and arcs, only; 
see Subsection \ref{subsect:special-cases}.
\item
Show that for arbitrary $t \in \R_+$ and for all
real analytic functions 
\bgl
m : [0,t] \nach \C\setminus\{0\}
 & \text{ \: and \: } & 
n : [0,t] \nach \R \,,
\egl
\noindent
the solutions $\faf$ and $\fbf$ of the 
equations \eqref{eq:dgl-faf}--\eqref{eq:dgl-fbf-anfbed1} are in $C_\AP(\R) + C_0(\R)$;
see Subsection \ref{subsect:general-case}.
\end{enumerate}


\subsection{Special Cases}
\label{subsect:special-cases}
In this subsection, we show that the parallel transports along 
straight lines and arc segments suffice to generate the almost periodic
and the vanishing-at-infinity functions.

\subsubsection{Straight Lines: Periodic}
Let $\gamma$ be a straight line along the $x$-axis, i.e., $m \ident 1$, $M \ident 0$ and $n \ident 0$. 
Then we have to solve $\ddot\faf + c^2 \faf = 0$ with $\faf(0) = 1$ and $\dot\faf(0) = 0$
and, similarly, for $\fbf$, getting
\bgl
\faf (t) & = & \cos ct \\
\fbf (t) & = & \I \sin ct \\
\egl\noindent
Tuning $t$ over $\R_+$, which corresponds to the different lengths straight edges may have,
we get all sine and cosine functions on $\R \ni c$ spanning a dense subspace in $C_\AP(\R)$,
hence
\blem
\label{lem:ap_in_algebra}
$C_\AP(\R)$ is contained in $C^\ast(\setabb^\ast \ptfktset)$.
\elem

\subsubsection{Spiral Arcs: Vanishing at $\infty$ plus Periodic}
Let $\gamma$ now be a path running with unit speed over (parts and/or multiples of) 
a spiral arc around the $z$-axis whose distance to the $z$-axis is denoted by $r$
and whose constant speed in $z$-direction by $\zspeed$. Moreover, we assume that
the $y$-component of $\gamma(0)$ vanishes and the spiral goes counterclockwise.
So, we have
\bgl
\gamma(t) 
  \breitrel\ident (x,y,z)(t)
  & = & \Bigl(r \cos \frac{\sqrt{1-\zspeed^2}}r \, t, 
                      r \sin \frac{\sqrt{1-\zspeed^2}}r \, t,
                      \zspeed \, t\Bigr) 
\egl\noindent
with $\betrag \zspeed < 1$, $r > 0$, and
\bgl[2ex]
m(t) & \ident & \dot x(t) - \I \dot y(t) 
      \breitrel= \sqrt{1-\zspeed^2} \: \e^{-\I\frac{\sqrt{1-\zspeed^2}}r \,t} \\
M(t) & \ident & \frac{\dot m(t)}{m(t)}
      \breitrel= -\I\frac{\sqrt{1-\zspeed^2}}r \\
n(t) & \ident & \dot z(t)
      \breitrel= \zspeed \,.
\egl\noindent
Equation \eqref{eq:dgl-fbf} now reads with $\paraml := \frac{\sqrt{1-\zspeed^2}}r > 0$
\bgl
\label{eq:spirale}
\ddot\fbf + \Bigl[\Bigl(c + \frac{\zspeed\paraml}2\Bigr)^2
                      + \frac{\paraml^2}4(1-\zspeed^2) \Bigr] \: \fbf 
 & = & 0\\
\egl\noindent
as it does for $\faf$. 
The initial conditions 
\eqref{eq:dgl-fbf-anfbed0} and
\eqref{eq:dgl-fbf-anfbed1} now imply, in particular,
\bgl
 \fbf(t) 
   & = & \frac{\I c \: \sqrt[4]{1-\zspeed^2}}%
               {\sqrt{\bigr(c + \frac{\zspeed\paraml}2\bigr)^2 + \frac{\paraml^2}4(1-\zspeed^2)}} 
	  \: \sin t \sqrt{\Bigr(c + \frac{\zspeed\paraml}2\Bigr)^2 + \frac{\paraml^2}4(1-\zspeed^2)} 
 \breitrel{=:} \fbf_{t,\zspeed,\paraml} (c)\,.
\egl\noindent
Note that the square roots are always nonzero.

\blem
For any $t \in \R$, $\betrag\zspeed <1$ and $\paraml > 0$, the functions 
\bgl
f_{t,\zspeed,\paraml} (c) 
  & := & \frac{\fbf_{t,\zspeed,\paraml}(c)}{\I \: \sqrt[4]{1-\zspeed^2}} - 
              \sin \Bigr(c + \frac{\zspeed\paraml}2\Bigr) t 
\egl
are smooth (also in the parameters) 
and vanish at infinity.
\elem
\bpf
This follows immediately from Lemma \ref{lem:asympt-period-arc}.
\qed
\epf
\noindent
Together with Lemma \ref{lem:ap_in_algebra}, this implies
\bcorr
\label{corr:arc_minus_sin_in_algebra}
Each $f_{t,\zspeed,\paraml}$ is contained in $C^\ast(\setabb^\ast \ptfktset)$.
\ecorr

\subsubsection{Separation of Points}

\blem
\label{lem:arcs_separating}
$\{f_{t,\zspeed,\paraml}\}_{t,\zspeed,\paraml}$ separates the points in $\R$
and vanishes nowhere.
\elem

\bpf
\bunum
\item
Let $c_1 < c_2$ be unseparable. This means
$f_{t,\zspeed,\paraml}(c_1) = f_{t,\zspeed,\paraml}(c_2)$
for all admissible parameters. Choose $\zspeed, \paraml$ with $c_1 + \frac{\zspeed\paraml}2 > 0$.
Observe that, considered as a function of $t$, the difference
$f_{t,\zspeed,\paraml}(c_1) - f_{t,\zspeed,\paraml}(c_2) \ident 0$ is a linear combination of
four sine functions with positive angular frequencies 
\bgl
& & c_1 + \frac{\zspeed\paraml}2
\,, \quad
\sqrt{\Bigr(c_1 + \frac{\zspeed\paraml}2\Bigr)^2 + \frac{\paraml^2}4(1-\zspeed^2)} 
\,, \quad \\
& & c_2 + \frac{\zspeed\paraml}2
\,, \quad
\sqrt{\Bigr(c_2 + \frac{\zspeed\paraml}2\Bigr)^2 + \frac{\paraml^2}4(1-\zspeed^2)}\,.
\egl
By assumption, the first frequency is smaller than any of the others.
Hence, by Lemma \ref{lem:sinus-lin-unabh}, the coefficient of the corresponding sine
function has to be zero, in contrast to the definition of $f_{t,\zspeed,\paraml}$
where the coefficient is one. Contradiction.
\item
If there was some $c$ with $f_{t,\zspeed,\paraml}(c) = 0$ for all parameters, 
we may use the same argumentation as above
to produce a contradiction.
\qed
\eunum
\epf

\bcorr
$C_0(\R)$ is contained in $C^\ast(\setabb^\ast \ptfktset)$.
\ecorr

\bpf
As the functions $f_{t,\zspeed,\paraml}$ are in $C_0(\R)$,
Lemma \ref{lem:arcs_separating} ensures that they span a dense 
subset of $C_0(\R)$ by the Stone-Weierstra\ss\ theorem.
On the other hand, by Corollary \ref{corr:arc_minus_sin_in_algebra},
each $f_{t,\zspeed,\paraml}$ is in 
the $C^\ast$-algebra $C^\ast(\setabb^\ast \ptfktset)$,
hence any function from $C_0(\R)$.
\qed
\epf

\subsubsection{Summary}

\bprop
\label{prop:C0_und_AP_sind_in_algebra}
$C_0(\R) + C_\AP(\R)$ is contained in $C^\ast(\setabb^\ast \ptfktset)$.
\eprop

\newcommand{\dummydot}{\bullet}
\newcommand{\beqa}{\begin{eqnarray}}
\newcommand{\eeqa}{\end{eqnarray}}
\DefThmUmgeb{Assumption}{Theorem}{\ZaehlerbisEbene}
\newcommand{\bass}{\begin{Assumption}}
\newcommand{\eass}{\end{Assumption}}
\newcommand{\propfakt}{\lambda}
\newcommand{\grenzmatrix}{K}
\newcommand{\restmatrix}{L}
\newcommand{\irgendeinefunktion}{s}
\newcommand{\su}{\mathfrak{su}}
\newcommand{\umbruch}[2][0ex]{#2{}\\&&\hspace*{#1}{}#2}

\subsection{General Case}
\label{subsect:general-case}
In this subsection, we are going to derive the inclusion relation
opposite to that of Proposition \ref{prop:C0_und_AP_sind_in_algebra}.
In other words, we have to show that the parallel transport along any 
given path is a linear combination of an almost periodic and a vanishing-at-infinity
function. Even more, we will see that periodicity, not just almost periodicity appears.

\subsubsection{Differential Equation to be Solved}
Let us now consider the following differential equation
\begin{eqnarray}
\label{eq:allg_dgl}
\ddot\faf + c^2 \faf
  & = & (\ffa + c \ffb)  \faf, 
\end{eqnarray}\noindent
together with the initial conditions
\begin{eqnarray}
\label{eq:allg_initial1}
\dot\faf(0) & = & \I c \ffd + \ffdd \\
\label{eq:allg_initial0}
    \faf(0) & = & \ffe . 
\end{eqnarray}\noindent
Here, $\ffa$ and $\ffb$ are real-analytic functions on some interval $[0,t]$.
Let us assume that both $\Im\ffa$ and $\Im\ffb$
are sign-conserving.%
\footnote{%
Note that a real-valued function $\varphi$ is 
called \df{sign-conserving} iff it is is
nonnegative or nonpositive. We define $\sgn\varphi$ to be $+1$ in the former
case and $-1$ in the latter one; if $\varphi \ident 0$, we may take either $+1$ or
$-1$. Of course, we have 
$(\sgn \varphi) \varphi = \betrag \varphi$ for any sign-conserving $\varphi$.}
As above, we may restrict ourselves to these
cases as we may decompose the paths, if necessary, such that the
respective functions $\Im\ffa$ and $\Im\ffb$ are sign-conserving along 
each single subpath. Indeed, this might shrink $\ptfktset$ further, but will
not change $C^\ast(\setabb^\ast\ptfktset)$, as above.
Finally, let $\ffd, \ffdd, \ffe$ be some fixed complex numbers,
and let $c$ be some real parameter. 
We are now interested in how $\faf(t)$ depends on $c$.

\subsubsection{General Solution}
Let us assume until Subsubsection \ref{subsubsect:first-factor} that $c>0$. 
We define on $[0,t]$ the constant function%
\beqa
\ffg & := & - \I \:\: \sgn (\Im\ffb) \: (\supnorm{\Im \ffa} + 1),
\nonumber
\eeqa
and let 
\beqa
\fff & := & -c^2 + (\ffa + c \ffb) - \ffg\,.
\nonumber
\eeqa

\bprop
\label{prop:olver-appl}
On $[0,t]$, the differential equation \eqref{eq:allg_dgl} has two real-analytic solutions%
\footnote{We choose the branches of the fractional powers of $\fff$ 
depending continuously on the
parameter $\tau\in[0,t]$ and on $c\in\R$,
such that, moreover,
$\sqrt\fff$ is the square of $\sqrt[4] \fff$.}
\begin{eqnarray}
\label{eq:fafpm-def}
\faf_\pm 
   & = & \frac{\e^{\pm \int_0^\dummydot \sqrt{\fff}}}{\sqrt[4] {\fff}} \: 
	 (1 + \varepsilon_\pm)\,.\nonumber
\end{eqnarray}
Here $\dummydot$ denotes the argument of $\faf_\pm$.
The error functions $\varepsilon_\pm : [0,t] \nach \C$ can be estimated by 
\beqa
\label{eq:absch-fehler-term-allg}
\supnorm{\varepsilon_\pm}, \: \Bigsupnorm{\frac{\dot\varepsilon_\pm}{\sqrt \fff}}
 & \leq & \e^{\int_0^t \betrag{\dot F}} - 1
\eeqa 
with
\beqa
F & := & \inv{\sqrt[4] \fff} \: \frac{\dd^2}{\dd^2 t} \Bigl(\inv{\sqrt[4] \fff}\Bigr) - \frac{g}{\sqrt \fff} \,.
\nonumber
\eeqa
\eprop
\blem
\label{lem:eigenschaften_fff}
Neither $\re \sqrt\fff$ nor $\Im\sqrt\fff$ nor $\fff$ have a zero on $[0,t]$.
Moreover,
\bglklein
\sup_{[0,t]} \betrag{\fff + c^2} 
 \breitrel\leq \frac{c^2}2 
 \breitrel\leq \inf_{[0,t]} \betrag\fff
\eglklein
for sufficiently large $c$.
\elem
\bpf
$\re\sqrt\fff$ or $\Im\sqrt\fff$ vanish iff 
$\Im\fff = 0$, i.e.,
\beqa
0 & = & \sgn (\Im\ffb) \: \Im\fff 
  \breitrel= \sgn (\Im\ffb) \: (\Im \ffa - \Im g + c \:\Im \ffb) \nonumber\\
  & = & \sgn (\Im\ffb) \: \Im \ffa + (\supnorm{\Im \ffa} + 1) + c \betrag{\Im \ffb} \nonumber
  \breitrel\geq 1\,. \nonumber 
\eeqa
The second assertion is trivial.
\qed
\epf

\bpf[Proposition \ref{prop:olver-appl}]
Since $\fff$ and $\ffg$ are real analytic on (some open set containing) $[0,t]$,
we may extend them to holomorphic functions (again denoted by $\fff$ and $\ffg$)
on some (simply connected) domain $D \teilmenge \C$ containing $[0,t]$.
Shrinking $D$, if necessary, we may assume that $\fff$ does not vanish on $D$.

Now, the proposition follows from Lemma \ref{lem:eigenschaften_fff} above 
and Theorem 11.1 in Section 6 of \cite{Olver}.  We only have to guarantee that there
exist points $a_1, a_2 \in D$, such that each $\tau \in [0,t]$ can be joined in $D$
with $a_1$ and $a_2$ by piecewise smooth arcs each having non-vanishing tangent vectors, 
such that
\bgl
\intsqrtf & := & \int \sqrt \fff,
\egl
has non-decreasing real part along these arcs for $a_1$ and non-increasing for
$a_2$.\ \cite{Olver} Assuming therefore, for the moment, $\re \sqrt \fff > 0$ on $[0,t]$, we see from
the definition of $\intsqrtf$ that
$\re\intsqrtf$ is indeed non-decreasing along the straight line from $a_1 := 0$ to $\tau \in [0,t]$,
but non-increasing along the straight line from $a_2 := t$ 
to $\tau \in [0,t]$. In the event of $\re \sqrt \fff < 0$,
simply exchange the r\^oles of $a_1$ and $a_2$, if necessary.
\qed
\epf

\subsubsection{Error-Term Estimate}

\newcommand{\nurfallendefkts}[1][1]{{\cal O}'_{#1}}
\newcommand{\fallendefkts}[1][1]{{\cal O}_{#1}}
\newcommand{\fallendefktspur}{{\cal O}_{c1}}
\newcommand{\dummyfkt}{h}
\newcommand{\compactum}{C}

\bdf
\label{def:abfall}
Let $\dummyfkt_c : [0,t] \nach \C$ be a function for each $c \geq 0$, and let $p \in \R$.
We say
\bgl
\text{$\{\dummyfkt_c\}$ is in $\fallendefkts[p]$\,.}
 & \aequ & \text{Each $\dummyfkt_c$ and $\Bigl\{c^p \, \bigsupnorm{\dummyfkt_c}\Bigr\}$ is bounded.} \\
\text{$\{\dummyfkt_c\}$ is in $\nurfallendefkts[p]$\,.}
 & \aequ & \textstyle \text{There is some compactum $\compactum$, such that} \\
 && \text{$\dummyfkt_c$ for $c \not\in C$ and $\Bigl\{c^p \, \bigsupnorm{\dummyfkt_c}\Bigr\}_{c\not\in C}$ is bounded.} \\
\egl
\edf\noindent
Naturally extending our notation, 
we simply write $\dummyfkt \in \fallendefkts[p]$,
if $\dummyfkt$ is a function that depends on $c$. Similarly, we extend it to $c$-depending constants 
(interpreted as constant functions on $[0,t]$).
And, we admit sets of functions for $\nurfallendefkts[p]$, even if they are defined and
bounded for
sufficiently large $c$ only.
\blem
We have $\dot F \in \fallendefkts$.
\elem
\bpf
Since
\bgl
\dot F & = & -\frac{8 \dddot\fff \fff^2 - 32 \ddot\fff \dot\fff \fff + 25 \dot\fff^3}{32\sqrt\fff^7} 
              + \frac{\ffg \fff}{2\sqrt \fff^3}\,,
\egl
we have, using Lemma \ref{lem:eigenschaften_fff}, 
\beqa
\supnorm F 
  & \leq & \frac{8 \supnorm{\dddot \fff} \supnorm \fff^2 + 32 \supnorm{\ddot \fff} \supnorm{\dot \fff} \supnorm \fff + 25 \supnorm{\dot \fff}^3}{2 \sqrt 2 \, c^7} 
                    + \frac{\sqrt 2 \betrag \ffg \supnorm{\dot \fff}}{c^3} \nonumber
\eeqa
for sufficiently large $c$.
The assertion now follows, since $\ffg$ is constant and the nominator of the first addend
can be estimated by a polynomial of degree $5$ in $c$:
\bgl[0.6ex]
\supnorm{\fff}      & \leq & c^2 + c \: \supnorm{\ffb} + \supnorm{\ffa - \ffg} \s
\supnorm{\dot\fff}  & \leq & \phantom{c^2 +{}} c \: \supnorm{\dot\ffb} + \supnorm{\dot\ffa} \,.
\egl
The estimates for higher derivatives are analogous to that for $\dot\fff$.
\qed
\epf
\noindent
Using the estimates from Proposition \ref{prop:olver-appl}, we get
\bcorr
\label{corr:bounded_variation_verschwindet_01}
Both $\varepsilon_\pm$ and $\frac{\dot\varepsilon_\pm}{\sqrt \fff}$ are in $\fallendefkts$.
\ecorr
\bpf
Use
\bglklein
c \, \Bigl(\e^{\int_0^t \betrag{\dot F}} - 1\Bigr)
 \breitrel\leq  c \, 
                \Bigl(\e^{\inv{c}
		        \sup_{c \geq 0} \bigl(c \, \supnorm{\dot F}\bigr) \, t} - 1\Bigr) 
 & \gegen & \sup_{c \geq 0} \bigl(c \, \supnorm{\dot F}\bigr) \, t
\eglklein
for $c \gegen \infty$, the continuity of $\dot F$ and \eqref{eq:absch-fehler-term-allg}.
\qed
\epf
\subsubsection{Initial Value Problem}

\blem
We have for sufficiently large $c$:
\bunum
\item
$\faf_+$ and $\faf_-$ are linear independent;
\item
$\faf_+$ and $\faf_-$ vanish nowhere.
\eunum
\elem
\bpf
According to Corollary \ref{corr:bounded_variation_verschwindet_01}, for sufficiently large $c$, 
we have $\supnorm{\varepsilon_\pm} < 1$. 
Now, $\faf_\pm$ can no longer vanish somewhere, since
$\fff$ and the exponential function are nowhere vanishing.
Assume next that $\faf_+$ and $\faf_-$ are linear dependent.
Then, 
$\faf_+(0) \faf_- = \faf_-(0) \faf_+$ on $[0,t]$, implying
\bgl
\e^{2\int_0^\dummydot  \sqrt{\fff}}
   & = & 
          \frac{1 + \varepsilon_+(0)}{1 + \varepsilon_-(0)} \: 
          \frac{1 + \varepsilon_-}{1 + \varepsilon_+}
\egl
Deriving this expression on $[0,t]$ yields
\bgl
2 \sqrt{\fff} \: \e^{2\int_0^\dummydot \sqrt{\fff}}
   & = & 
          \frac{1 + \varepsilon_+(0)}{1 + \varepsilon_-(0)} \: 
          \frac{\dot\varepsilon_- (1 + \varepsilon_+) - \dot\varepsilon_+ (1 + \varepsilon_-)}{(1 + \varepsilon_+)^2}
\egl
Both equations together give
\beqa
\label{eq:fehler_linunabh}
          \frac{\dot\varepsilon_-}{2 \sqrt{\fff}} (1 + \varepsilon_+) 
	- \frac{\dot\varepsilon_+}{2 \sqrt{\fff}} (1 + \varepsilon_-)
 & = & 
         (1 + \varepsilon_-)(1 + \varepsilon_+)\,.
\eeqa
Now, the estimates of 
Corollary \ref{corr:bounded_variation_verschwindet_01}
show that
\bgl
\Bigsupnorm{\frac{\dot\varepsilon_\pm}{\sqrt \fff}}
 & \text{ \: and \: } & 
\supnorm{\varepsilon_\pm}
\egl
are smaller than $\inv4$ for sufficiently large $c$. Then, the norm of the left hand side
of \eqref{eq:fehler_linunabh} is smaller than $2 \cdot \inv2 \cdot \inv4 \cdot \frac54 = \frac5{16}$,
but that of the right hand side is larger than $\frac34 \cdot \frac34 = \frac9{16}$.
This is a contradiction.
\qed
\epf

\bcorr
For sufficiently large $c$, the solution $\faf$ of the initial value problem
\eqref{eq:allg_dgl}--\eqref{eq:allg_initial0}
equals
$\propfakt_+ \faf_+ + \propfakt_- \faf_-$
with coefficients $\propfakt_\pm$ that depend continuously on $c$.
\ecorr
%

\subsubsection{Factorization}
To get the full estimate, consider
\beqa
\label{eq:fund_solutions}
\propfakt_\pm \faf_\pm
 & = & \frac{\propfakt_\pm}{\sqrt[4]{\fff(0)}} \cdot 
        \frac{\sqrt[4]{\fff(0)}}{\sqrt[4]{\fff}} \cdot
        \e^{\pm \int_0^\dummydot \sqrt{\fff}} \cdot
	 (1 + \varepsilon_\pm)\,.
\eeqa\noindent
We will next decompose all four factors appropriately into a sum of a $c$-uniformly bounded function 
and a continuous $\fallendefkts$ function. (The fourth factor is trivial, of course.)
The oscillating term will emerge from the exponential.

\subsubsection{Second Factor}
\blem
\label{lem:factor2}
We have 
\beqa
\frac{\sqrt[4]{\fff(0)}}{\sqrt[4]{\fff}} - 1 & \in & \fallendefkts \,. \nonumber
\eeqa
\elem
\bpf
Use Lemma \ref{lem:eigenschaften_fff} to get
\beqa
\Bigbetrag{\frac{\fff(0)}{\fff} -1}
 & \ident & \frac{\bigbetrag{\ffa(0) - \ffa + c (\ffb(0) - \ffb)}}%
                 {\bigbetrag{\fff}}
 \breitrel\leq  4 \: \frac{\supnorm{\ffa} + c \: \supnorm{\ffb}}%
                 {c^2} \nonumber
\eeqa
for large $c$, hence $\frac{f(0)}f - 1$ is in $\fallendefkts$.
As the root branches are chosen continuously (zero is not passed), the assertion follows.
\qed
\epf

\subsubsection{Third Factor}
As $\Im \sqrt\fff$ never vanishes, we may 
assume for simplicity that $\Im\sqrt\fff$ is positive on $[0,t]$.
\blem
\label{lem:sqrtfff-abschaetzung}
We have
\beqa
\sqrt\fff - \I \Bigl(c - \frac{\ffb}2 \Bigr) \breitrel\in \fallendefkts \,. \nonumber
\eeqa
\elem
\bpf
Since $\Im\fff$ is assumed to be positive, we have 
\bglklein[1.5ex]
 \bigbetrag{\sqrt{\fff} + \I \bigl(c - \frac{\ffb}2 \bigr)} 
 & \geq & \bigbetrag{c + \Im \sqrt{\fff}  - \re\frac{\ffb}2} 
 \breitrel\geq c - \inv2 \supnorm{\ffb}
\eglklein
for sufficiently large $c$. Hence, for such $c$
\bgl
\Bigbetrag{\sqrt{\fff} - \I \Bigl(c - \frac{\ffb}2 \Bigr)}
 & = & \frac{\bigbetrag{\fff + \bigl(c - \frac{\ffb}2 \bigr)^2}}
            {\bigbetrag{\sqrt{\fff} + \I \bigl(c - \frac{\ffb}2 \bigr)}} 
 \breitrel\leq \frac{\bigsupnorm{\ffa + \inv4 \ffb^2 - \ffg}}{c - \inv2 \supnorm{\ffb}} \,. \nonumber 
\egl
The continuity of the term under investigation gives the assertion.
\qed
\epf

\bcorr
\label{corr:exponential-abschaetzung}
Recalling $\charact_{c} (\tau) = \e^{\I c \tau}$, we have
\bgl
\e^{\pm \int_0^\dummydot \sqrt{\fff}}
   - \e^{\mp \frac\I2 \int_0^\dummydot \ffb} \: \charact_{\pm c}
 & \in & \fallendefkts \nonumber \,.
\egl
\ecorr
\bpf
For brevity, we take the upper signs only.
Using $\supnorm{\e^\irgendeinefunktion - 1} \leq \e^{\supnorm\irgendeinefunktion} - 1$
and $\supnorm{\e^\irgendeinefunktion} \leq \e^{\supnorm\irgendeinefunktion}$,
we have
\beqa
\bigsupnorm{\e^{\int_0^\dummydot \sqrt{\fff}}
   - \e^{- \frac\I2 \int_0^\dummydot \ffb} \:\charact_{c}}
 & = & \bigsupnorm{\e^{\I \int_0^\dummydot (c - \frac{\ffb}2)} 
            \: \bigl(\e^{\int_0^\dummydot {(\sqrt{\fff} - \I(c - \frac{\ffb}2)})} - 1 \bigr)}
\nonumber \\
 & \leq & \bigsupnorm{\e^{\I \int_0^\dummydot (c - \frac{\ffb}2)}} 
            \bigl(\e^{\supnorm{ \int_0^\dummydot {(\sqrt{\fff} - \I(c - \frac{\ffb}2)})}} - 1 \bigr)
\nonumber \\
 & \leq & \e^{\einhalb \supnorm{\ffb} t} 
            \bigl(\e^{\supnorm{\sqrt{\fff} - \I(c - \frac{\ffb}2)} t} - 1 \bigr)\,.
\nonumber
\eeqa
The assertion follows from Lemma \ref{lem:sqrtfff-abschaetzung}.
\qed
\epf

\subsubsection{First Factor}
\label{subsubsect:first-factor}
\blem
\label{lem:faf_0-abschaetzung}
We have
\bgl
\faf_\pm(0) \sqrt[4]{\fff(0)} - 1 \breitrel\in \fallendefkts & \text{ \: and \: } & 
\frac{\dot\faf_\pm(0)}{\I c} \sqrt[4]{\fff(0)} \mp 1 \breitrel\in \nurfallendefkts \nonumber \,.
\egl

\elem
\bpf
\bunum
\item
The first assertion follows from
\bgl
\faf_\pm(0) \sqrt[4]{\fff(0)} & = & 1 + \varepsilon_\pm
\egl
and Corollary \ref{corr:bounded_variation_verschwindet_01}.
\item
For the second part, first
observe that
$c \, \bigl(\frac{\sqrt\fff}{\I c} - 1\bigr)$ is bounded 
on each compact set not containing $c = 0$.
In fact, Lemma \ref{lem:sqrtfff-abschaetzung} implies
\bgl[2ex]
\Bigbetrag{\frac{\sqrt\fff}{\I c} - 1} 
 & \ident & \Bigbetrag{\frac{\sqrt\fff - \I (c - \inv2\ffb) - \frac\I2\ffb}{\I c}} \s
 & \leq & \frac{\supnorm{\sqrt\fff - \I (c - \inv2\ffb)} +\inv2\supnorm\ffb}{c} \,.
\egl\noindent
Next,
observe
\bgl
\dot\faf_\pm
   & = & \frac{\e^{\pm \int_0^\dummydot \sqrt{\fff}}}{\sqrt[4] {\fff}} \: 
	 \Bigl[\dot \varepsilon_\pm
	        -  \bigl(1 + \varepsilon_\pm\bigr)
                    \Bigl(\frac{\dot\fff}{4\fff} \mp \sqrt{\fff}\Bigr) \Bigr] \nonumber
\egl
giving
\bgl
\Biggl[\frac{\dot\faf_\pm}{\I c} \sqrt[4]{\fff} \mp 1\Biggr] \Bigeinschr0
 & = & \Biggl[\frac{\dot \varepsilon_\pm}{\sqrt{\fff}} \: \frac{\sqrt{\fff}}{\I c}
	 -  \varepsilon_\pm
                    \Bigl(\inv{\I c}\frac{\dot\fff}{4\fff} \mp \frac{\sqrt{\fff}}{\I c}\Bigr) 
	 -  \inv{\I c}\frac{\dot\fff}{4\fff}  {}\pm{} \Bigr(\frac{\sqrt{\fff}}{\I c} - 1\Bigr) \Biggr]\Bigeinschr0\,.
\egl
Now the assertion follows from \eqref{eq:absch-fehler-term-allg},
from $\frac{\dot\fff}{\fff} \in \fallendefkts$ and from
$\frac{\sqrt\fff}{\I c} - 1 \in \nurfallendefkts$
above.
\qed
\eunum
\epf

\blem
\label{lem:factor3}
We have
\beqa
\frac{\propfakt_\pm}{\sqrt[4]{\fff(0)}} - \frac{\ffe \pm \ffd}2 
 & \in & \nurfallendefkts \,. \nonumber
\eeqa
\elem

\bpf
From the initial value problem follows that 
\bgl
\I c \ffd + \ffdd & = & \dot\faf(0) \breitrel= \propfakt_+ \dot\faf_+(0) + \propfakt_- \dot\faf_-(0) \\
     \ffe & = &     \faf(0) \breitrel= \propfakt_+ \faf_+(0) + \propfakt_- \faf_-(0),
\egl
or, equivalently (for $c \neq 0$),
\bgl
\bpm \ffd \\ \ffe \epm + 
\inv{\I c} \bpm \ffdd \\ 0 \epm 
  \breitrel\ident \inv{\sqrt[4]{\fff(0)}} \:
                    (\grenzmatrix + \restmatrix) \: \bpm \propfakt_+ \\ \propfakt_- \epm
\egl
with
\bgl
\grenzmatrix := \bpm 1 & -1 \\ 1 & 1 \epm
& \text{ \: and \: } & 
\restmatrix := \bpm \frac{\dot\faf_+(0)}{\I c} \sqrt[4]{\fff(0)} - 1 & \frac{\dot\faf_-(0)}{\I c} \sqrt[4]{\fff(0)} + 1 
                                            \\ \faf_+(0) \sqrt[4]{\fff(0)} - 1 & \faf_-(0) \sqrt[4]{\fff(0)} - 1 \epm \,.
\egl

By Lemma \ref{lem:faf_0-abschaetzung}, we have 
$\grenzmatrix^{-1} \restmatrix \in \nurfallendefkts$. 
So $\norm{\grenzmatrix^{-1} \restmatrix} < 1$ 
for large $c$, whence
$\grenzmatrix + \restmatrix$ is invertible with
\bglklein
(\grenzmatrix + \restmatrix)^{-1} - \grenzmatrix^{-1}
 & = & (1 + \grenzmatrix^{-1} \restmatrix)^{-1} \: \grenzmatrix^{-1}- \grenzmatrix^{-1}
 \breitrel= \sum_{i=1}^\infty (-\grenzmatrix^{-1} \restmatrix)^{i} \grenzmatrix^{-1}
\eglklein
in $\nurfallendefkts$, since
\bglklein[1ex]
\norm{\sum_{i=1}^\infty (\grenzmatrix^{-1} \restmatrix)^{i} \grenzmatrix^{-1}}
 \breitrel= \norm{\grenzmatrix^{-1}} \frac{\norm{\grenzmatrix^{-1} \restmatrix}}{1 - \norm{\grenzmatrix^{-1} \restmatrix}} \,.
\eglklein
Using $\grenzmatrix^{-1} = \einhalb\grenzmatrix^T$, we get the assertion from
\beqa
&&
\hspace*{-9em} \inv{\sqrt[4]{\fff(0)}} \bpm \propfakt_+ \\ \propfakt_- \epm - \inv2 \bpm \phantom{-}\ffd + \ffe \\ -\ffd + \ffe \epm 
\nonumber \\[-1.586ex]
  & = & \Bigl[(\grenzmatrix + \restmatrix)^{-1} - \einhalb \grenzmatrix^T\Bigr] 
          \Biggl[\bpm \ffd \\ \ffe \epm + \inv{\I c} \bpm \ffdd \\ 0 \epm \Biggr] 
	  + \inv{\I c} \bpm \ph-\ffdd \\ -\ffdd \epm\,. \hspace*{-5em}
\nonumber
\eeqa
\qed
\epf

\subsubsection{Final Result}

\bprop
The solution $\faf$ of the differential equation \eqref{eq:allg_dgl} with
initial conditions \eqref{eq:allg_initial1} and \eqref{eq:allg_initial0}
equals 
\bgl
\grenzfaf 
 & := & \frac{\ffe + \ffd}2 \cdot 
          \e^{- \frac\I2 \int_0^\dummydot \ffb} \: \charact_{c} +
        \frac{\ffe - \ffd}2 \cdot 
          \e^{+ \frac\I2 \int_0^\dummydot \ffb} \: \charact_{-c}
\egl\noindent
plus a bounded real-analytic function $\restfaf$. The latter function depends on $c$
in such a way that 
\bgl
\text{$\{\supnorm{c \, \restfaf} \}_{c\in\R}$ is bounded.}
\egl
\eprop
\bpf
Defining $\restfaf := \faf - \grenzfaf$, we get 
$\restfaf \in \nurfallendefkts$ for $c \geq 0$ from the decomposition \eqref{eq:fund_solutions}
together with \eqref{eq:absch-fehler-term-allg}, Lemma \ref{lem:factor2},
Corollary \ref{corr:exponential-abschaetzung} and Lemma \ref{lem:factor3}.
Moreover, 
the case $c \leq 0$ can be reduced to the case of $c \geq 0$ by replacing $c$, $\ffd$ and $\ffb$
by $-c$, $-\ffb$ and $-\ffb$, respectively, which is a transformation that leaves 
$\grenzfaf$ invariant. Therefore, we even know that
$\{\supnorm{c \, \restfaf} \}_{c\in\R\setminus\compactum}$ is bounded for some compactum $\compactum$.
The remaining statement on $\compactum$, however, is obvious, as \eqref{eq:allg_dgl} is a
linear differential equation and as $[0,t] \kreuz C$ is compact.
Finally, as $\faf$ and $\grenzfaf$ are analytic, also $\restfaf$ is analytic for each $c$. 
\qed
\epf

\noindent
As we know from Subsection \ref{subsect:parallel-transport_dgl} that
any element in $\setabb^\ast \ptfktset$
is (possibly, up to an overall constant prefactor)
a solution of \eqref{eq:allg_dgl}, \eqref{eq:allg_initial1}
and \eqref{eq:allg_initial0} for appropriate coefficients, we get

\bprop
\label{prop:algebra_enthaelt_AP+C_0}
$\setabb^\ast \ptfktset$ is contained in $C_0(\R) + C_\AP(\R)$.
\eprop

\noindent
Propositions \ref{prop:C0_und_AP_sind_in_algebra} and \ref{prop:algebra_enthaelt_AP+C_0},
together with Corollary \ref{corr:AP+C_0_ist_C*-alg}
and the final remark in Subsection \ref{subsect:selection-of-paths}, now imply
\bthm
\label{thm:algebra_gleich_AP+C_0}
$C^\ast(\setabb^\ast \ptfktset) = C^\ast(\setabb^\ast \alg_2)$ equals $C_0(\R) \dirvsum C_\AP(\R)$.
\ethm
From the derivation in Subsection \ref{subsect:special-cases}
we see that already the straight lines and the spiral arcs suffice to 
span a dense subset in $C_0(\R) \dirvsum C_\AP(\R)$.

\brem
As communicated to us by Martin Bojowald, Tim Koslowski has independently claimed 
\cite{d84} that the parallel transports depend asymptotically periodic on $c$ 
and, moreover, can approximate any asymptotically almost periodic function arbitrarily well. 
In our paper, we have given a rigorous proof for these facts.
\erem

\subsection{Configuration Space for Homogeneous Isotropic $k = 0$ LQC}
Let us summarize the results derived in Subsections \ref{subsect:special-cases}
and \ref{subsect:general-case}:
\bthm
Let $\alg_2$ be the $C^\ast$-subalgebra of $\linf(\A)$
generated by the parallel transport matrix functions along all piecewise analytic 
paths in $M = \R^3$. Moreover, let $\setabb : \R \nach \A$ be the embedding
$c \auf c \invzush$ with $\invzush = \tau_1 \dd x + \tau_2 \dd y + \tau_3 \dd z$
being homogeneous isotropic.
Define $\alg_1$ to be the $C^\ast$-subalgebra of $\boundfkt(\R)$ generated by 
the restriction algebra $\setabb^\ast \alg_2$.

Then $\alg_1$ equals the vector space sum 
\bgl
C_0 (\R) \dirvsum C_\AP(\R) 
 & \teilmenge & \boundfkt(\R)
\egl
of the algebra of continuous functions on $\R$ vanishing at infinity, plus
the algebra of almost periodic functions on $\R$.
Its spectrum is given by the $\iota$-twisted sum \cite{paper46}
\bgl
\quer\R & := & \R \disjunion \rb \,,
\egl
where $\iota : \R \nach \rb$ is the natural mapping.
\ethm
Recall from \cite{paper46} that the $\iota$-twisted sum topology is generated by the following
three types of open sets:
\bgl
\begin{array}{lcrclcl}
        && \voffen & \!\!\!\sqcup\!\!\! & \leeremenge 
        && \text{with open $\voffen \teilmenge \R$} \\
        && \compl\vkomp & \!\!\!\sqcup\!\!\! & \rb
        && \text{with compact $\vkomp \teilmenge \R$} \\
        && \iota^{-1}(W) & \!\!\!\sqcup\!\!\! & W
        && \text{with open $W \teilmenge \rb$\,.}
\end{array}
\egl
\noindent
As the almost periodic functions generate the topology on $\rb$ via Gelfand transform,
the third type of sets can also be replaced by
\bgl
\begin{array}{lcrclcl}
        && f^{-1}(U) & \!\!\!\sqcup\!\!\! & \widetilde f^{-1}(U) 
        && \text{with open $U \teilmenge \C$ and $f \in C_\AP(\R)$,}
\end{array}
\egl\noindent
where we used the relation 
$\gelf f \circ \iota = f$ (see Lemma \ref{lem:gelf-iota=elalg}).

\brem
Without touching the mathematical content of the theorem, one can, of course, argue that
$\malg_1$ above is not the physically correct configuration space of homogeneous isotropic loop quantum
cosmology. In fact, the Bohr compactification has been very successfully used
in LQC, and one could even 
say that one can get the desired embedding property by restricting the 
algebra $\alg_2$ of full loop quantum gravity to, say, piecewise linear paths. This option has
been studied by Engle \cite{d85}. We, however, do not think that this is the best way.
In fact, loop quantum gravity should comprise all different types of cosmologies. So we should not
form our full theory after a single reduced theory as then we may be given non-embedding 
results for other symmetric models. Instead, if any, the symmetric models shall be ruled by the full theory.
\erem


\section{Conclusions}
We conclude with some comments on possible extensions of the present paper.
\bunum
\item
First of all, one can further investigate the properties of the 
solution of the differential
equation, in particular, its full expansion into powers of $\inv c$. 
Of course, this includes a proof that
the solution $\faf$ is real-analytic at $\pm\infty$. 
Brunnemann and Koslowski have proceeded in that direction.%
\footnote{After the first version of our article 
had been put on the {\tt arxiv}, Brunnemann and Koslowski 
completed their preprint \cite{d87}.}
They have derived a recursion equation
for the coefficients of the power series and are going to establish the 
necessary estimates for all orders in $\inv c$.
Moreover, they describe $\quer\setabb$ explicitly in terms of spin networks
and discuss the implementation of symmetries further on the quantum level of the full 
gravitational theory.

\item
Then, one should determine the behaviour of parallel transports for
the spherical ($k=1$) and the hyperbolic ($k=-1$) homogeneous isotropic universes.
We expect completely analogous behaviour if one replaces straight lines by geodesics
and $\invzush$ by the respective (up to gauge transforms) homogeneous isotropic 
connection. 
\item
Next, one should investigate the homogeneous, but anisotropic case. Here, 
we already know from \cite{paper28} that generically the parallel transports do not
depend almost periodically on $c$ (for $k=0$). Even more, they are ``at least as
non-almost periodic'' as for the corresponding isotropic case. This can easily be seen
as the isotropic connections form a diagonal line $\R$ in the set of anisotropic connections
forming $\R^3$. The detailed analysis, however, will be more sophisticated, as the 
nice structure of the differential equation \eqref{eq:dgl2_a} for $a$, where the (w.r.t.\ $c$) 
leading coefficient of $a$ is constantly $c^2$ is now quadratic in $c_1, c_2, c_3$ though, but
path-depending:
\begin{eqnarray}
\label{eq:dgl-a-aniso}
\ddot a + (c_1^2 \dot x^2 + c_2^2 \dot y^2 + c_3^2 \dot z^2) \,a
 & = & \I \Bigl(c_3 \ddot z 
               - c_3 \dot z \frac{c_1 \ddot x - \I c_2 \ddot y}{c_1 \dot x - \I c_2 \dot y}
	       \Bigr) \,a +
        \frac{c_1 \ddot x - \I c_2 \ddot y}{c_1 \dot x - \I c_2 \dot y} \,\dot a  \,.
\end{eqnarray}
One easily sees that \eqref{eq:dgl-a-aniso} reduces to \eqref{eq:dgl2_a} if $c_1 = c_2 = c_3 = c$
and if the path $\gamma$ is parametrized w.r.t.\ to the arc length (as then
$\norm{\dot\gamma}^2 \ident \dot x^2 + \dot y^2 + \dot z^2 = 1$).
\item
Our choice that the reduced algebra is given by that of the full theory, has a further advantage:
We now may impose symmetries successively. Thus, we expect that the respective embedding properties 
show a simple functorial behaviour. 
\item
Finally, a big step towards a fully quantized model will be the selection of a measure on 
$\quer\R = \R \disjunion \rb$. Until now, the Haar measure on the Bohr compactification
served as the canonical measure to give the Hilbert space. Now, observe that 
still $\rb$ is a subset of $\quer\R$, but it is no longer a dense subset. Thus, the justification
of again singling out the Haar measure is difficult; for recent developments see the updates
section below. Probably, a full explanation will only be possible after
investigating the full phase space structure of the reduced theory. Nevertheless, naively, one
could take any measure on $\rb$ and any measure on $\R$, and then ``add'' them.
The standard Lebesgue measure on $\R$, however, seems not so appropriate as 
the asymptotically vanishing part of the symmetric spin-network functions is of order $\inv c$,
hence usually not integrable.
\eunum


\section{Updates}
Recently, after the first preprint version of our article had appeared at the {\tt arxiv}, 
the issue 
of measures has been discussed from several points of views.
\bunum
\item
Ashtekar and Campiglia \cite{d89} studied the reduced Weyl algebra
in the Bianchi I case. They obtained that this algebra has a 
unique cyclic representation that is invariant w.r.t.\ residual 
diffeomorphisms. This beautiful result is a direct analogy to the Stone-von Neumann
like theorem for full
loop quantum gravity.\ \cite{paper18}
One should, nevertheless, emphasize that 
the reduced Weyl algebra, as used in \cite{d89}, 
took only straight lines into account. Therefore,
it is not surprising that Ashtekar and Campiglia obtained $L^2(\rb,\mu_\Haar)$
as the Hilbert space distinguished by diffeomorphism invariance. 
\item
Engle \cite{d88}, in contrast, incorporated all analytic paths.
At the same time, as motivated by \cite{d89}, he 
opted for the Haar measure on the $\rb$-part of $\quer\R$,
giving the $\R$-part measure zero. This way, he balanced between the 
Ashtekar-Campiglia result and the embeddability of $\R$ into $\quer\R$.
Consequently, Engle was able to re-implement the Bojowald-Kastrup
idea of LQC states as symmetric LQG states. 
\item
Hanusch \cite{max-diss}, very recently, showed that the Haar measure 
on $\rb$, considered as a measure on $\quer\R$, is the only normalized
regular Borel measure that is invariant under the action of $\R$ on $\quer\R$
that is induced by the standard action on $\R$ by left translations.
This is a strong hint that the Ashtekar-Campiglia result might be extended
also to our framework.
\item
Beyond that, using projective structures, Hanusch \cite{d111}
constructed a family of measures
convexly combining the Haar measure on $\rb$ as used by Engle and a normalized measure
on $\R$ that is induced from the Lebesgue measure via an appropriate homeomorphism
between $(0,1)$ and $\R$. This way, Hanusch obtained three different types of auxiliary
Hilbert spaces (with two of them isomorphic). However, the implementation on the
level of states is still open.
\eunum
Finally, one should note that all the approaches described above obtained measures on
$\rb$ or $\quer\R$, both seen as spectral compactification of the classical reduced
configuration space $\R$. In other words, first the symmetry has been implemented and then
the system got quantized. Very recently, Hanusch \cite{d110} studied the other order: he
first lifted group actions from the full classical configuration space $\A$ to
the quantum configuration space $\Ab$ and then studied the corresponding
invariant connections. It turned out, that typically the reduced quantum
configuration space $\Abinv$ is strictly larger than the quantized
reduced configuration space $\R$.


\section{Acknowledgements}
The author thanks Johannes Brunnemann and Maximilian Hanusch
for numerous discussions and many helpful comments on 
a draft of the present article.
In particular, the author is very much obliged to Maximilian Hanusch 
as he detected an error in the spectrum stated in the first version of the article.
Moreover, the author gratefully acknowledges 
discussions with
Thomas Tonev concerning asymptotically almost periodic functions.
The author has been supported by the Emmy-Noether-Programm of
the Deutsche Forschungsgemeinschaft under grant FL~622/1-1.

\anhangengl


\section{Separation Property for Smooth Connections}
\label{app:separate}
First of all, assume that we are working in one of the following cases
$G_\omega$, $G_\infty$, $G_k$, and $G_{\mathrm{PL}}$ (see Subsubsection \ref{subsubsect:techn-param-lqg}).
Consider some path
$\gamma:[-1,1] \nach M$ which is at least $C^k$-smooth and 
covered by an appropriate trivialization of the bundle, 
and denote the subpath $\gamma \einschr{[0,t]}$ by $\gamma_t$.
The parallel transports for a connection $A$ along 
these $\gamma_t$ are given by
\bgl
\ddfrac{}t h_A(\gamma_t) = - A(\dot\gamma_t) \: h_A(\gamma_t)
 & \text{ \: with \: } &  h_A(\gamma_0) = \EINS.
\egl\noindent
So we can reconstruct $A(\dot\gamma_t)$ for any $t$ out of the parallel transports.

Now, for each nonzero tangent vector $X$ in $M$, there is a path $\gamma$ in the game as above
with 
$\dot \gamma(0) = X$.
Therefore, even the connection $A$ itself can be reconstructed uniquely from 
the respective parallel transport matrix functions.
In particular, these functions separate any two distinct connections in $\A$.


\section{Closedness Property for Almost Periodic Functions}
For completeness, in this appendix, we present a direct proof that the 
asymptotically almost periodic functions form a $C^\ast$-subalgebra
of $\boundfkt(\R)$. For this, we first show that the algebras of
almost periodic functions (being a $C^\ast$-subalgebra of $\boundfkt(\R)$)
and that of continuous functions vanishing at 
infinity (being an ideal in $\boundfkt(\R)$) 
have trivial intersection. Actually, that the sum of both algebras is
closed, hence a $C^\ast$-subalgebra, is already guaranteed by general 
arguments \cite{Murphy} or by connecting 
Propositions \ref{prop:C0_und_AP_sind_in_algebra} and \ref{prop:algebra_enthaelt_AP+C_0}.
Here, however, we will derive this result directly, even for all non-compact
LCA groups%
\footnote{An LCA group is a locally compact abelian topological Hausdorff group.}.

\blem
\label{lem:norm-add-sup-AP}
Let $G$ be a connected noncompact LCA
group.
Then we have
\bgl
\supnorm{f} & \leq & \supnorm{h+f}
\egl
for all $f \in C_\AP(G)$ and all $h \in C_0(G)$.
\elem
Recall \cite{m15,m14}
that a function $f \in \boundfkt(G)$ is almost periodic 
iff its translate set \mbox{$\{L^\ast_g f \mid g \in G\}$} is relatively compact;
in other words, iff each sequence $(L^\ast_{g_i} f)$ contains a 
subsequence that converges in $\boundfkt(G)$.
Moreover, observe \cite{RudinFour} 
that an LCA group $G$ has some open (and closed) subgroup $G_1$ being the
product of some $\R^n$ and some compact group $C$. If $G$ is connected, $G$
even equals $G_1$, and if $G$ is additionally noncompact, then $n > 0$. 
Therefore we may assume that $G$ equals $\R \kreuz H$ for some topological
group $H$. Also, by $\pi : G \nach \R$, we denote the canonical projection.

\bpf
Let $\varepsilon > 0$.
\bunum
\item
Choose some compact $K \teilmenge G$, such that $\betrag h < \varepsilon$ 
on the complement of $K$.

Then we have 
$\betrag f \leq \betrag{h + f} + \betrag h < \supnorm{h + f} + \varepsilon$
on $\compl K$,
hence 
\bgl
\supnorm{h + f}
 & > & \sup_{\compl K} \betrag f - \varepsilon\,.
\egl

\item
Choose now some $g \in G$, such that $\betrag{f(g)} > \supnorm f - \varepsilon$.
\bunum
\item
As $f$ is continuous, we may w.l.o.g.\ assume that $\pi(g) \neq 0$.
\item
Let first be $g \in K$.
As $\pi(K) \teilmenge \R$ is compact, we find some $r\in\N_+$, such that
$\pi(K) \teilmenge [-r,r]$. As $\betrag{\pi(g^s)} = \betrag{s \pi(g)} > r$
for integer $\betrag{s} > S := \ogkl{r/\betrag{\pi(g)}}$, we have $s g \notin K$ for 
$\betrag s > S$.

As $f$ is almost periodic, there is a strictly increasing sequence $(n_k)$, such that
$(L^\ast_{n_k S g} f)$ is converging, hence Cauchy. Selecting
appropriate $k_1 > k_2$, we find 
\bgl
\supnorm{L^\ast_{nSg} f - f} 
 & = & \supnorm{L^\ast_{n_{k_1} Sg} f - L^\ast_{n_{k_2} Sg}  f} 
 \breitrel< \varepsilon
\egl
with $n := n_{k_1}-n_{k_2}$, implying
\bgl
\betrag{f([nS+1]g)} 
 & \ident & \betrag{L^\ast_{nSg} f(g)} 
 \breitrel> \betrag{f(g)} - \varepsilon 
 \breitrel> \supnorm f - 2\varepsilon \,.
\egl
As $nS+1 > S$, we have $[nS+1]g \not\in K$, hence
\bgl
\sup_{\compl K} \betrag{f} 
 & \geq & \betrag{f([nr+1]g)} 
 \breitrel> \supnorm f - 2\varepsilon
\egl
\item
If now $g \not\in K$, then we know directly from the choice of $g$ that 
\bgl
 \sup_{\compl K} \betrag f
 & \geq & \betrag{f(g)} 
 \breitrel> \supnorm f - \varepsilon \,.
\egl
\eunum
Altogether, we get

\bgl
\supnorm{h + f}
 & > & \sup_{\compl K} \betrag f - \varepsilon
 \breitrel >   \supnorm{f} - 3 \varepsilon \,.
\egl

\eunum
As $\varepsilon$ was arbitrary, we get the proof.
\qed
\epf

\bcorr
\label{corr:AP+C_0_ist_C*-alg}
Let $G$ be a connected noncompact LCA group. Moreover,
let $\alg_0 \teilmenge C_0(G)$ 
and $\alg_1 \teilmenge C_\AP(G)$ be $C^\ast$-subalgebras (hence of
$\boundfkt(G)$ as well).

Then $\alg_0 \cap \alg_1 = \NULL$, and $\alg_0 + \alg_1$ is closed in
$\boundfkt(G)$.
\ecorr

\bpf
To see $\alg_0 \cap \alg_1 = \NULL$, 
assume $f \in \alg_0 \cap \alg_1$. Then Lemma \ref{lem:norm-add-sup-AP}
applied to $h := -f$ gives $\supnorm f \leq \supnorm{f + (-f)} = 0$,
hence $f = 0$.

To prove the closedness of $\alg_0 + \alg_1$, consider a Cauchy sequence
$(h_n + f_n) \teilmenge \alg_0 + \alg_1 \teilmenge \boundfkt(G)$.
Since $\supnorm{f_n - f_m} \leq \supnorm{h_n - h_m + f_n - f_m}$,
again by Lemma \ref{lem:norm-add-sup-AP},
also $(f_n)$ is a Cauchy sequence converging to some $f \in \alg_1$.
Consequently, $(h_n)$ converges, whence so does $(h_n + f_n)$.
\qed
\epf


\section{Sine Function Properties}
\blem
\label{lem:asympt-period-arc}
For $\paramd,\parame \in \R$, $\parame \neq 0$, the function $f : \R \nach \R$, defined by
\bgl
f(c) & := & \frac{c+\paramd}{\sqrt{c^2 + \parame^2}} \sin {\sqrt{c^2 + \parame^2}}
              - \sin c
\egl
vanishes at infinity and is smooth in $c, \paramd, \parame$.
\elem
\bpf
Smoothness is obvious. For the other claim, 
consider $\fktg(c) := \sqrt{1 + \frac{\parame^2}{c^2}} - 1$ for
$c \neq 0$.
Obviously, $\fktg \gegen 0$ for $\betrag c \gegen \infty$, 
but also $\betrag c \fktg \gegen 0$.
Since, for nonzero $c$,
\bgl[2ex]
f(c) 
  & = & 
             \Biggl[\frac{1+\frac{\paramd}c}{1 + \fktg} \cos (c\fktg) - 1 \Biggr] \sin c 
              + \Biggl[\frac{1+\frac{\paramd}c}{1 + \fktg} \sin (c\fktg) \Biggr] \cos c \,,
\egl
the claim follows as both brackets vanish 
for $\betrag c \gegen \infty$, whereas $\sin c$ and $\cos c$ remain bounded. 
\qed
\epf
\noindent
For completeness, we derive finally the well-known

\blem
\label{lem:sinus-lin-unabh}
$\{\sin_\lambda\}_{\lambda>0} \teilmenge \boundfkt(\R)$, with
$\sin_\lambda(x) := \sin \lambda x$, is linear independent.
\elem
\bpf
Let $\sum_{k = 1}^n a_k \sin_{\lambda_k} = 0$ with $0 < \lambda_1 < \ldots < \lambda_k$ and 
$a_k \in \C$. Taking the $2N$-th derivative, we get
$\sum_{k = 1}^n a_k \lambda^{2N}_k \sin_{\lambda_k} = 0$,
hence
\bgl
a_n
 \breitrel\ident a_n \sin_{\lambda_n} \Bigl(\frac{\pi}{2\lambda_n}\Bigr)
 & = & \sum_{k = 1}^{n-1} a_k \Bigl(\frac{\lambda_k}{\lambda_n}\Bigr)^{2N} 
            \sin_{\lambda_k} \Bigl(\frac{\pi}{2\lambda_n}\Bigr)
\egl
for all $N$.
But, as $\lambda_k < \lambda_n$ for $k<n$, the right-hand side 
goes to zero for $N \gegen \infty$,
whence $a_n = 0$. The proof follows by induction.
\qed
\epf


\end{document}